\documentclass[a4wide,]{article}

\pdfoutput=1

\usepackage{etex}
\usepackage{tikz}
\usetikzlibrary{patterns}

\usepackage{acronym}
\usepackage{algorithm}
\usepackage{algorithmicx}
\usepackage{amsmath}
\usepackage{amssymb}
\usepackage{array}
\usepackage{authblk}
\usepackage{booktabs} 
\usepackage{boxedminipage}
\usepackage{color}
\usepackage{colortbl}
\usepackage{crop}
\usepackage{enumitem}
\usepackage{epic}
\usepackage{epsfig}
\usepackage{fancyhdr}
\usepackage{fancyvrb}
\usepackage{float}
\usepackage[left=120pt, right=120pt]{geometry}
\usepackage{graphics}
\usepackage{graphicx}
\usepackage{ifthen}
\usepackage{listings}
\usepackage{longtable}
\usepackage{makeidx}
\usepackage{moreverb}
\usepackage{multicol}
\usepackage{multirow}
\usepackage{pgfplots}
\usepackage{pslatex}
\usepackage{rotating}
\usepackage{setspace}
\usepackage{soul}
\usepackage{subfig}
\usepackage[dotinlabels]{titletoc}
\usepackage{ulem}
\usepackage{url}

\newcolumntype{I}{!{\vrule width 1.5pt}}
\newlength\savedwidth
\newcommand\whline{\noalign{\global\savedwidth\arrayrulewidth
                            \global\arrayrulewidth 1.5pt}%
           \hline
           \noalign{\global\arrayrulewidth\savedwidth}}

\newboolean{IsWide}
\setboolean{IsWide}{true}

\newcommand{\FlaTwoByTwo}[4]{
\left( 
\begin{array}{c I c}
#1 & #2 \\ \whline
#3 & #4 
\end{array} 
\right)
}

\newcommand{\FlaThreeByThreeTL}[9]{
\left( 
\begin{array}{c | c I c}
#1 & #2 & #3 \\ \hline
#4 & #5 & #6 \\ \whline
#7 & #8 & #9
\end{array} 
\right) 
}

\newcommand{\FlaThreeByThreeBR}[9]{
\left( 
\begin{array}{c I c | c}
#1 & #2 & #3 \\ \whline
#4 & #5 & #6 \\ \hline
#7 & #8 & #9
\end{array} 
\right)
}

\newcommand{\FlaPartition}[2]{
\ifthenelse{\boolean{IsWide}}{{\bf partition } \hspace{-1em} #1 \hspace{-1em} #2}
{{\bf partition } \+ \\ #1 \+ \\ #2 \- \-}
}

\newcommand{\FlaRepartition}[2]{
\ifthenelse{\boolean{IsWide}}{{\bf repartition } \hspace{-1em} #1 \hspace{-1em} #2}
{{\bf repartition } \+ \\ #1 \+ \\ #2 \- \-}
}

\newcommand{\FlaContinue}[1]{
\ifthenelse{\boolean{IsWide}}{{\bf continue with } #1
}
{{\bf continue with } \+ \\ #1 \-
}
}

\newcommand{\FlaStartComputeShorter}{
\setlength{\unitlength}{0.5in}
\begin{picture}(3,0.01)
\put(0,0){\line(1,0){3}}
\put(0,0.01){\line(1,0){3}}
\end{picture} 
}

\newcommand{\FlaEndComputeShorter}{
\setlength{\unitlength}{0.5in}
\begin{picture}(3,0.01)
\put(0,0){\line(1,0){3}} 
\put(0,0.01){\line(1,0){3}} 
\end{picture} 
}

\newcommand{\operation}{ [ D, E, F, \ldots ] \becomes {\rm op}( A, B, C, D, \ldots ) }
\newcommand{\routinename}{ [ D, E, F, \ldots ] \becomes {\rm op}( A, B, C, D, \ldots ) }
\newcommand{\routinecost}{ X }
\newcommand{\precondition}{ Q }
\newcommand{\postcondition}{ R }
\newcommand{\invariant}{ P }
\newcommand{\costinv}{ \  }

\newcommand{\guard}{ R }
\newcommand{\partitionings}{
\begin{minipage}{3in}
$ S_I $
\end{minipage}
}
\newcommand{\partitionsizes}{ \hspace{ 3.25in} }

\newcommand{\blocksize}{1}

\newcommand{\repartitionings}{
\begin{minipage}[t]{3in}
\ \\
\ \\
\ \\
\end{minipage}
}

\newcommand{\repartitionsizes}{ \hspace{ 3.25in} }
\newcommand{\moveboundaries}{
\begin{minipage}[t]{3in}
\ \\
\ \\
\ \\
\end{minipage}
}
\newcommand{\beforeupdate}{
$ \QBefore $
}
\newcommand{\afterupdate}{
$ \QAfter $
}
\newcommand{\update}{%
\begin{minipage}[t]{4in}
$ S_U $
\end{minipage}
}

\newcommand{\resetsteps}{
\renewcommand{\blocksize}{1}
\renewcommand{\operation}{ [ D, E, F, \ldots ] \becomes {\rm op}( A, B, C, D, \ldots ) }
\renewcommand{\routinename}{ [ D, E, F, \ldots ] \becomes {\rm op}( A, B, C, D, \ldots ) }
\renewcommand{\routinecost}{ 0 }
\renewcommand{\precondition}{ \PPre }
\renewcommand{\postcondition}{ \PPost }
\renewcommand{\invariant}{ \PInv }
\renewcommand{\costinv}{ \  }
\renewcommand{\guard}{ G }
\renewcommand{\partitionings}{ %
\begin{minipage}[t]{3in}
\ \\
\end{minipage}
}
\renewcommand{\partitionsizes}{ \hspace{ 3.25in} }
\renewcommand{\repartitionings}{%
\begin{minipage}[t]{3in}
\ \\
\end{minipage}
}
\renewcommand{\repartitionsizes}{ \hspace{ 3.25in} }
\renewcommand{\moveboundaries}{%
\begin{minipage}[t]{3in}
\ \\
\end{minipage}
}
\renewcommand{\beforeupdate}{
\QBefore
}
\renewcommand{\afterupdate}{
\QAfter
}
\renewcommand{\update}{
$ S_U $
}
}

\newcommand{\WSguard}{
$ \guard $
}

\newcommand{\WSpartitionNarrow}{
\begin{minipage}[t]{2.0in}
\begin{tabbing}
id \= id \= \kill
{\bf Partition} 
\partitionings \+ \\
{\bf where } 
\begin{minipage}[t]{1.5in}
\partitionsizes 
\end{minipage}
\end{tabbing}
\end{minipage}
}

\newcommand{\WSrepartition}{
\begin{minipage}[t]{3in}
\ifthenelse{ \equal{\blocksize}{1} }{}
{%
\ifthenelse{ \equal{\blocksize}{blank} }{~}
{{\bf Determine block size $ \blocksize $}} \\
}
{\bf Repartition}
\begin{tabbing}
in \= in \= \+ \kill
\repartitionings \+ \\
{\bf where } \hspace*{-2ex} \repartitionsizes 
\end{tabbing}
\end{minipage}
}

\newcommand{\WSrepartitionNarrow}{
\begin{minipage}[t]{2.05in}
\ifthenelse{ \equal{\blocksize}{1} }{}
{%
\ifthenelse{ \equal{\blocksize}{blank} }{~}
{{\bf Determine block size $ \blocksize $}} \\
}
{\bf Repartition}
\begin{tabbing}
i \= i \= \+ \kill
\repartitionings \+ \\
{\bf where }
\begin{minipage}[t]{1.5in}
\repartitionsizes
\end{minipage}
\end{tabbing}
\end{minipage}
}

\newcommand{\WSmoveboundaryNarrow}{
\begin{minipage}[t]{2.05in}
{\bf Continue with}
\begin{tabbing}
i \= \+ \kill
\moveboundaries 
\end{tabbing}
\end{minipage}
}

\newcommand{\WSupdate}{
\update
}

\newcommand{\FlaAlgorithmNarrow}{
\begin{center}
\begin{tabular}{|l |} \hline
{\bf Algorithm:} $\routinename$ 
\\ \whline
{\WSpartitionNarrow} \\
{\bf while} \WSguard { \bf do} \\
\ \hspace{0.0in} \WSrepartitionNarrow \\
{\hspace{0.0in} \FlaStartComputeShorter} \\
{\hspace{0.0in} \WSupdate} \\
{\hspace{0.0in} \FlaEndComputeShorter} \\
{\ \hspace{0.0in} \WSmoveboundaryNarrow} \\
{{\bf endwhile}} \\ \hline
\end{tabular}
\end{center}
}

\newcommand{ \PPre }{ P_{\it pre} }
\newcommand{ \PPost }{ P_{\it post} }
\newcommand{ \PInv }{ P_{\it inv} }

\newcommand{ \becomes }{:=}
\newcommand{ \QBefore }{ P_{\it before} }
\newcommand{ \QAfter }{ P_{\it after} }

\pgfplotsset{compat=newest}
\pgfplotscreateplotcyclelist{journal}{%
blue,every mark/.append style={fill=blue!80!black},mark=square*\\%
red,densely dashed,every mark/.append style={solid,fill=red!80!black},mark=o\\%
brown!80!black,every mark/.append style={fill=brown!80!black},mark=diamond\\%
green,every mark/.append style={fill=green!80!black},mark=diamond*\\%
cyan,densely dashed,every mark/.append style={solid,fill=cyan!80!black},mark=*\\%
magenta,densely dashed,every mark/.append style={solid,fill=magenta!80!black},mark=square*\\%
green,densely dashed,every mark/.append style={solid,fill=red!80!black},mark=o\\%
} 

\newcommand\lstsetbylang[1]{
  \lstset{
    language=#1,
    showstringspaces=false,
    formfeed=\newpage,
    tabsize=4,
    commentstyle=\itshape,
    basicstyle=\ttfamily\footnotesize, 
    morekeywords={in,endif,endfor,end,begin,function},
    keywordstyle=\bf,
    commentstyle=\it,
    aboveskip=-0.3em,
    belowskip=-0.5em
  }
}

\floatstyle{boxed}
\floatname{code}{Figure}
\newfloat{code}{btp}{lop}

\makeatletter
\let\c@code\c@figure
\makeatother

\newcommand\Tab[1]{{Table~\ref{tab:#1}}}
\newcommand\Fig[1]{{Fig.~\ref{fig:#1}}}
\newcommand\Sec[1]{{Section~\ref{sec:#1}}}
\newcommand\Code[1]{{Fig.~\ref{code:#1}}}
\newcommand\ie{{\it i.e.}}
\newcommand\eg{{\it e.g.}}

\newcommand{\tr}[1]{{#1}^{\mathrm{T}}}

\newcommand{\Chol}[1]{\mbox{\sc Chol}(#1)}

\newcommand{\triu}[1]{\mbox{\sc triu}(#1)}

\newcommand{\changeout}[1]{%
\sout{#1}%
}

\renewcommand{\changeout}[1]{}

\newcommand{\bfA}{\mbox{\boldmath $A$}}

\newcommand{\bfP}{\mbox{\boldmath $P$}}

\acrodef{AMD}{Approximate Minimum Degree}
\acrodef{API}{Application Programming Interface}

\acrodef{BFS}{Breadth First Search}
\acrodef{BLAS}{Basic Linear Algebra Subprogram}
\acrodef{BLACS}{Basic Linear Algebra Communication Subprograms}
\acrodef{CG}{Conjugate Gradient}

\acrodef{CSR}{Compressed Sparse Row}
\acrodef{CSC}{Compressed Sparse Column} 

\acrodef{DAG}{Directed Acyclic Graph}
\acrodef{DAGUE}{Directed Acyclic Graph Unified Environment}
\acrodef{DFS}{Depth First Search}
\acrodef{DLA}{Dense Linear Algebra}

\acrodef{DOF}{Degree Of Freedom}
\acrodefplural{DOF}{Degrees of Freedom}

\acrodef{FEB}{Full Empty Bit}
\acrodef{FEM}{Finite Element Method}
\acrodef{FE}{Finite Element}
\acrodef{FIFO}{First-In First-Out}
\acrodef{FLAME}{Formal Linear Algebra Methods Environment}
\acrodef{FLASH}{Formal Linear Algebra Scalable Hierarchical}
\acrodef{FLOP}{FLoating Point Operation} 
\acrodef{FSB}{Front-Side Bus}

\acrodef{GAS}{Global Address Space}
\acrodef{GPU}{Graphic Processing Unit}
\acrodef{HPC}{High Performance Computing}
\acrodef{HPX}{High Performance ParalleX}

\acrodef{LAPACK}{Linear Algebra PACKage}
\acrodef{LGPL}{Lesser General Public License}
\acrodef{LIFO}{Last-In First-Out}
\acrodef{LINAL}{LINear ALgebra}
\acrodef{MAGMA}{Matrix Algebra on GPU and Multicore Architectures}
\acrodef{MKL}{Math Kernel Library}
\acrodef{MUMPS}{MUltifrontal Massively Parallel sparse direct Solver}
\acrodef{MPI}{Message Passing Interface}

\acrodef{ND}{Nested Dissection}
\acrodef{NSF}{National Science Foundation}
\acrodef{NUMA}{Non-Uniform Memory Access}
\acrodef{OOC}{Out-Of-Core}
\acrodef{PAPI}{Performance Application Programming Interface}
\acrodef{PaRSEC}{Parallel Runtime Scheduling and Execution Controller}
\acrodef{PCG}{Preconditioned Conjugate Gradient}
\acrodef{PLASMA}{Parallel Linear Algebra for Scalable Multi-core Architectures}
\acrodef{PME}{Partitioned Matrix Expression}
\acrodef{TACC}{Texas Advanced Computing Center}
\acrodef{QPI}{QuickPath Interconnect}
\acrodef{QUARK}{QUeuing And Runtime for Kernels}
\acrodef{ScaLAPACK}{Scalable Linear Algebra PACKage}
\acrodef{SIMD}{Single Instruction Multiple Data}
\acrodef{SIMT}{Single Instruction Multiple Threads}
\acrodef{SMP}{Symmetric Multi-Processing}
\acrodef{SPD}{Symmetric Positive Definite}
\acrodef{SSD}{Solid State Device}
\acrodef{RCM}{Reverse Cuthill McKee}
\acrodef{TLB}{Translation Look-aside Buffer}
\acrodef{TBB}{Threading Building Blocks}
\acrodef{UHM}{Unassembled HyperMatrix}
\acrodefplural{UHM}{Unassembled HyperMatrices}

\def\keywords{
  {\bfseries\textit{Keywords}---\,\relax
}}

\newcommand{\kj}[2]{{{#1}}{}}

\newcommand{\figsize}{0.3}
\newcommand{\figup}{0cm}

\newcommand{\legendsize}{1.4in}
\newcommand{\legendfont}{\scriptsize}
\newcommand{\axisfont}{\footnotesize}
\newcommand{\graphwidth}{3.8in}
\newcommand{\graphheight}{2.8in}

\newcommand\blfootnote[1]{%
  \begingroup
  \renewcommand\thefootnote{}\footnote{#1}%
  \addtocounter{footnote}{-1}%
  \endgroup
}

\title{Task Parallel Incomplete Cholesky Factorization using 2D
  Partitioned-Block Layout} 
\author[1]{Kyungjoo~Kim}
\author[1]{Sivasankaran~Rajamanickam}
\author[1]{George~Stelle}
\author[1]{H.~Carter~Edwards}
\author[1]{Stephen~L.~Olivier}
\affil[1]{Center for Computing Research, Sandia National Laboratories}
\affil[ ]{\{kyukim,srajama,gwstell,hcedwar,slolivi\}@sandia.gov}

\date{}

\begin{document}            

\begin{titlepage}
  \maketitle



\begin{abstract}
  We introduce a task-parallel algorithm for sparse
  incomplete Cholesky factorization that utilizes a
  2D 
  sparse partitioned-block layout of a matrix.
  Our factorization algorithm follows the idea of 
  {\it algorithms-by-blocks} by using the block layout. 
  The algorithm-by-blocks approach induces a task graph for the factorization.
  These tasks are inter-related to each other through their data
  dependences in the factorization algorithm.
  To process the tasks on various manycore architectures in a
  portable manner, we also present a portable tasking API that
  incorporates different tasking backends and device-specific
  features using an open-source framework for manycore
  platforms \ie, Kokkos.
  A performance evaluation is presented on both Intel
  Sandybridge and Xeon Phi platforms for matrices from the University
  of Florida sparse matrix collection to illustrate merits of the
  proposed task-based factorization. Experimental results demonstrate
  that our task-parallel implementation delivers about 26.6x speedup 
  (geometric mean) over single-threaded incomplete Cholesky-by-blocks
  and 19.2x speedup over serial Cholesky performance which does not
  carry tasking overhead using 56 threads on the Intel Xeon Phi processor
  for sparse matrices arising from various application problems.
\end{abstract}


  \blfootnote{
    \noindent This technical report is a preprint of a paper intended for
    publication in a journal or proceedings. Since changes may be made
    before publication, this preprint is made available with the
    understanding that anyone wanting to cite or reproduce it ascertains
    that no published version in journal or proceedings exists. \\

    The code described in this paper is publicly available at
    \url{https://github.com/trilinos/Trilinos/tree/master/packages/shylu/tacho}.\\
    
    Sandia is a multiprogram laboratory operated by Sandia
    Corporation, a Lockheed Martin Company, for the U.S. Department of
    Energy under contract DE-AC04-94-AL85000.  
  }

\end{titlepage}



\acresetall

\section{Introduction}

Incomplete Cholesky factorization is effectively used
for preconditioned iterative methods
to solve large-scale \ac{SPD} linear systems.
Computing incomplete factorizations scalably in shared-memory systems 
is an open problem for both multicore and manycore
architectures, because incomplete factorizations are
characterized by irregular data access patterns, frequent
synchronizations, 
and dependences that limit the available parallelism when expressed in a
data-parallel manner. 
First, incomplete factorizations, by definition, are much
more sparse than their 
counterparts, the complete factorizations.
This sparsity precludes the use of any \ac{DLA} operations such as
the \ac{BLAS} kernels and results in a sparse data access pattern 
that is very irregular in a traditional incomplete factorization
algorithm.
By traditional, we refer to the incomplete factorization algorithm
(\eg, left-looking and right-looking variants)
that is
implemented using a compressed sparse row/column 
format.
Second, a traditional parallel incomplete factorization uses an ordering
technique of rows or/and columns to expose some parallelism.
\kj{However, the conventional sparse factorization algorithms}
   {While this can be done recursively, algorithms such as these}
   still suffer from synchronization bottlenecks for rows or columns
   that cannot be factored in parallel. 
Third, when the only available parallelization option is 
using a simple \texttt{parallel for},
the traditional incomplete factorization
cannot be expressed efficiently. In general, a matrix is reordered to
explore parallelism and the substructure resulting from the reordering
phase is more suitable for {\it task-parallel} algorithms,
which require means of expression as such.

We propose a parallel incomplete sparse factorization algorithm and 
its implementation targeting multicore and manycore architectures, called
{\it Tacho}. 
In particular, we focus on task-parallel sparse
level(k) incomplete Cholesky factorization.\footnote{\kj{The level(k)
  incomplete factorization determines the location of additional
  nonzero factors, called $fills$, based on the sparsity pattern of a
  matrix. Initially, all nonzero entries of the matrix are set with a
  level $0$. Then, a fill is created with an increasing level and restricted by the
  threshold $k$ during Gaussian elimination.}{}}
Our approach is based on a class of algorithms, called
{\it algorithms-by-blocks}, originating from parallel out-of-core
\ac{DLA} algorithms~\cite{Gunter:2005:poc}. 
This class of algorithms has been adopted for
asynchronous thread-parallel execution in \ac{DLA}
libraries~\cite{Buttari:2009:tiles,Chan:2007:supermatrix,Quintana:2009:blocks}.
\kj{%
  Applying this style of algorithms to sparse matrix factorization involve
  several challenges in handling the irregular data structure of
  sparse matrices and blocking strategies that can expose the
  parallelism.}{}

In \ac{DLA}, several
\acp{API}~\cite{Gustavson:1999:packed,Low:2004:flash,Valsalam:2002:matrix}
are proposed to facilitate algorithms-by-blocks. These \acp{API}
are primarily developed to improve data locality by changing the
standard columnwise storage format to a recursive block storage
format. However, no attempt has been made to use a similar 2D
block layout on sparse matrices for shared-memory factorizations. The sparse
linear algebra community has considered block-based layouts for
simple kernels such as sparse matrix vector multiply~\cite{siva_sc}
or sparse matrix-matrix multiply~\cite{aydin_csb}.
Instead, 1D data layouts are often used for high performance
computing libraries that utilize factorizations~\cite{Hysom:2001:ilu}. 
The use of 1D partitions can severely limit parallelism when a sparse
matrix has a large bandwidth and it incurs synchronization
bottlenecks. On the contrary, the 2D block layout 
based on \ac{ND} ordering is more suitable to expose fine-grained
task parallelism and better load balance as it can create a number of
tasks that can be concurrently executed.
This enables the asynchronous task parallelism in the block level rather than in
a row level. The usage of such a layout also regularizes the data access
with respect to the blocks.
\kj{The subblocking approach is mostly applied to sparse direct 
factorizations for computing dense (supernodal)
blocks~\cite{Hogg:2010:dag,Irony:2004:sparse_chol,Kim:2014:sparse,Rothberg:1994:cholesky}.}{}
To the best of our knowledge,
this 2D sparse partitioned-block layout, where {\it the blocks
of a sparse matrix are themselves sparse}, has not been explored
for complex kernels like sparse factorizations in shared-memory
architectures. For brevity, we will use the term 2D layout
or 2D block matrix for a two-dimensional sparse partitioned-block
layout and a matrix that uses that layout respectively. 
From an implementation perspective, the 2D block, a light-weight
object that describes a rectangular computing region on a sparse matrix, becomes an
entry of a 2D block matrix (matrix of blocks). Note 
that the hierarchical representation of our block sparse matrix
does not need to repack data associated with a block;
instead, the block points to the base matrix with
appropriate meta data (partition information) specifying the rectangular
region. Further performance improvement can be achieved by
repacking the corresponding data of blocks. However, repacking may
carry additional overhead and the cost might be
significant considering the light workload of incomplete
factorization.


By applying algorithms-by-blocks on the 2D layout, a problem is
reformulated in terms of block matrix computations;
blocks become a computing unit and operations among blocks become
tasks. Then, resulting tasks are scheduled, potentially {\it
  out-of-order}, to compute resources after satisfying 
task dependences. In short, the depedences expressed through the \ac{API} 
define a partial order of possible task executions that the runtime system 
maps to available threads and the particular system architecture.
This approach yields a clear separation of concerns
by decoupling algebraic structure from runtime task scheduling.

We have implemented the task-parallel Cholesky factorization by extending the 
open-source Kokkos library~\cite{Kokkos:2014:JPDC} with a portable
tasking interface and using it for the factorization.
Our implementation is available in the ShyLU
 package~\cite{Rajamanickam:2012:shylu} of the Trilinos library.
The Kokkos library provides a high-level programming abstraction pursuing
portable performance on various manycore architectures.
We have extended it to include a portable interface for task parallelism.
Through the interface, developers write an application code once and the code is
portable to heterogeneous device environments with device-specific
programming models. Currently, our extensions include backends
for Pthreads and Qthreads~\cite{qthreads} to schedule task parallelism on host
devices, \eg, IBM \textsc{Power} series, Intel Xeon multicore and Intel Xeon Phi
manycore processors. Kokkos already provides support for data parallelism 
on the GPU, and we are on-track to develop a GPU backend for the
Kokkos task interface in the coming year.
Key features of the new interface include futures and dependences to enable 
general task \acp{DAG},
and non-blocking semantics to accommodate devices such as GPUs.

The main contributions of this paper include:
\begin{itemize}
\item 
  a high-level matrix abstraction for 2D sparse partitioned-block matrices
  that facilitates task parallelism with dependences
  using future references;
\item 
  a new task-parallel implementation for sparse level(k) 
  incomplete Cholesky factorization that utilizes 2D layouts;
\item 
  a portable tasking interface and its implementation,
  designed to support different tasking backends for different hardware
  features and limitations;
\item 
  performance evaluation for several test problems 
  that shows our task-parallel factorization-by-blocks
  delivers scalable and portable parallel
  performance on an Intel Sandybridge processor and an Intel Xeon Phi
  coprocessor.
\end{itemize}
The rest of the paper is organized as follows. \Sec{kokkos-task} 
describes our extensions of the Kokkos library 
to support task parallelism. \Sec{algo} explains sparse level(k) incomplete
Cholesky-by-blocks. A performance evaluation is presented in
\Sec{result}. Some related work and the conclusions of our work are preseneted 
in \Sec{related-work} and \Sec{conclusion} respectively.

\section{Kokkos portable tasking API}
\label{sec:kokkos-task}


The programming model chosen for Tacho 
is an extension of Kokkos~\cite{Kokkos:2014:JPDC} to support
dependency-driven task-parallel execution. 
Kokkos has been developed to address the challenge of
performance portability across manycore architectures;
\eg, multicore CPU, Intel Xeon Phi, and NVIDIA GPU. 
Until recently, Kokkos was supported only for data parallelism.
However, our extensions enable the specification of computational
tasks together with the dependence relationships between them. 
These tasks and dependences form an implicit \ac{DAG}
that is scheduled by a run time system on behalf of the
application. 

\subsection{Abstraction}


A Kokkos task is created with a C++ {\it functor} (body of work)
to execute and an optional number of {\it dependences}.
Dependences are defined by handles to other tasks that must complete before
the task scheduler will execute the newly created task.
Upon successful creation of a task, a {\it future} is returned.
A Kokkos future is the handle to a task that may be used to
denote inter-task dependences, probe for task completion status, or 
obtain the return value of a task.

A Kokkos {\it execution policy} defines how and where ``bodies of work''
will execute in parallel.
For the task interface, all run time task management occurs
through a {\it task execution policy}.
This policy is responsible for the creation, destruction,
scheduling, and execution of a group of related tasks.
Dependent tasks must be members of the same task execution policy
so that their dependences can be enforced by the policy.

\begin{code}[tb!]
  \lstsetbylang{C++}
  \begin{center}
    \begin{lstlisting}[mathescape]
// using the Kokkos namespace
void foo() {
  using Space = /* where to execute */ ;

  // the policy is defined on a specific execution space
  // multiple policy objects on different execution spaces are allowed 
  TaskPolicy<Space> policy ;

  // FunctorX,Y,Z are C++ classes containing tasks' code
  Future<Space> f_x = policy.create( FunctorX() );
  Future<Space> f_y = policy.create( FunctorY() );
  Future<Space> f_z = policy.create( FunctorZ() );
  policy.add_dependence( f_z , f_x ); // f_z depends on f_x
  policy.add_dependence( f_z , f_y ); // f_z depends on f_y
  policy.spawn( f_z ); // FunctorZ is now waiting on FunctorX and FunctorY
                       // to complete execution
  policy.spawn( f_x ); // may immediately execute
  policy.spawn( f_y ); // may immediately execute
  wait( policy ); // wait for all tasks to complete
}
  \end{lstlisting}
  \end{center}
  \caption{Simple example of using a Kokkos task execution policy
  to create tasks, introduce dependences, spawn tasks for execution,
  and wait for a group of tasks to complete.
  }
  \label{code:TaskPolicy}
\end{code}

We succinctly illustrate how a user creates a task execution policy
and tasks with dependences in \Code{TaskPolicy}.
In this example, three tasks are implemented with C++ classes:
\texttt{FunctorX}, \texttt{FunctorY}, and \texttt{FunctorZ}.
The policy's \texttt{create} function allocates a task with the
given functor implementation, but does not schedule the task.
The \texttt{add\_dependence} function introduces a dependence
of the first task upon the second task; \ie, the first
task is not allowed to execute until the second task completes.
The \texttt{spawn} function schedules the task for execution.
If there are no dependences, the scheduled task may immediately
execute, perhaps even completing before the \texttt{spawn}
function returns.
Finally, the \texttt{wait} function is called to wait for all
ready tasks owned by the policy to complete, including tasks
that become ready in the course of executing other tasks
owned by the policy.



Properties of Kokkos tasks and data are derived from goals of both productivity 
and performance portability. Support for task DAG execution based on 
dependences and futures allow significant flexibility in the types of 
algorithms that can be expressed and the amount of parallelism that can 
be exposed.
Tasks are non-preemptive and non-blocking because some architectures
targeted by Kokkos; \eg, tasks on GPUs cannot support blocking.

Kokkos data is expressed in the form of multidimensional arrays, called {\it views}.
A Kokkos view defines where allocated data resides
(\eg, CPU vs. GPU memory) and the {\it layout} of that
multidimensional array data.
Layout is polymorphic with respect to the execution architecture.
For example, on CPUs the default layout is row major (array of structures)
and on GPUs the default layout is column major (structure of arrays).
The view abstraction and API allows an architecture-appropriate
layout to be introduced into user code without requiring any modification
of that code.

\subsection{Implementation}


We support two tasking runtimes as the task execution policy for Kokkos.
A basic task execution policy using a Pthreads
thread-pool to execute tasks on shared memory multicore and manycore platforms. 
We also provide a task execution policy using the
Qthreads~\cite{qthreads} lightweight threading library for both
task scheduling and execution.
The Qthreads library is optimized for efficient and scalable
node-level execution on CPU-like multicore and manycore architectures.
Qthreads offers fast context switching between tasks and
software-implemented \ac{FEB} synchronizations, inspired
by the Tera MTA / Cray XMT architecture~\cite{mta}, that map especially
well to futures.
We plan to develop a task execution policy for GPU architectures within
the next year.

A task execution policy's scheduler has three primary responsibilities.
First, it tracks which tasks are ready for execution and which tasks
are waiting for inter-task dependences to be satisfied.
Second, it selects and executes ready tasks on available cores.
Third, as tasks complete it updates their associated dependent tasks
to a ready state.
The goal of a scheduler is to carry out these responsibilities
as efficiently and thread-scalably as possible.
Note that application code written using the Kokkos API does not need to be changed to
benefit from new or improved backend implementations.

\section{Sparse incomplete factorization}
\label{sec:algo}

This section describes task-based level(k) incomplete
Cholesky 
factorization using the 2D block layout. The factorization method
in Tacho consists
of symbolic factorization and numeric factorization.
Symbolic factorization (\Sec{algo:symb})
results in a 2D sparse partitioned-block matrix 
(\Sec{algo:twod}).
Once the sparsity pattern of the Cholesky factors is determined,
Cholesky-by-blocks numeric
factorization (\Sec{algo:byblocks}) generates tasks with dependences according
to the data flow of the factorization process and computes factors via the portable
tasking \ac{API}.  


\subsection{Symbolic factorization}
\label{sec:algo:symb}

\begin{algorithm}[tbh]
  \begin{algorithmic}[1]
    \State Compute the ND ordering 
    \State Reorder the matrix with the ND ordering 
    \State Prune $t$ levels of the ND tree to control minimum block sizes
    \State Find the level-$k$ fill
    \State Construct 2D block matrix based on the ND tree 
  \end{algorithmic}
  \caption{Symbolic factorization}
  \label{alg:symb_fact}
\end{algorithm}

Algorithm~\ref{alg:symb_fact} describes our symbolic factorization phase.
As a first step, we use a \acf{ND}
ordering~\cite{George:1973:nd} 
from the Scotch~\cite{scotch:6.0:manual} library to expose a high degree of
concurrency during the factorization.  The algorithm recursively separates a
matrix into two subproblems, 
providing a tree hierarchy. 
In addition to improving the concurrency, the \ac{ND} ordering also reduces
the amount of fill (zeros turning into non-zeros during the factorization),
which corresponds to the amount of work in the factorization.
To keep the same amount of work during scaling studies, we generate
the same number of levels of the \ac{ND} tree but optionally prune
the tree to allow enough work for each task.
Next, we determine the location of potential fill up to
the given fill-level using graph analysis.
Our 
implementation of the symbolic factorization follows the
algorithm proposed by Hysom and Pothen~\cite{Hysom:2002:ilu}.
Using the \ac{ND} tree, we construct a
2D sparse partitioned-block matrix (a sparse matrix of sparse
matrices).
We devote the next subsection to this last step.


\subsection{2D sparse partitioned-block matrix}
\label{sec:algo:twod}

\renewcommand{\figsize}{0.8}
\begin{figure}
  \begin{center}
    \subfloat[Natural ordering]{
      \includegraphics[scale=\figsize]{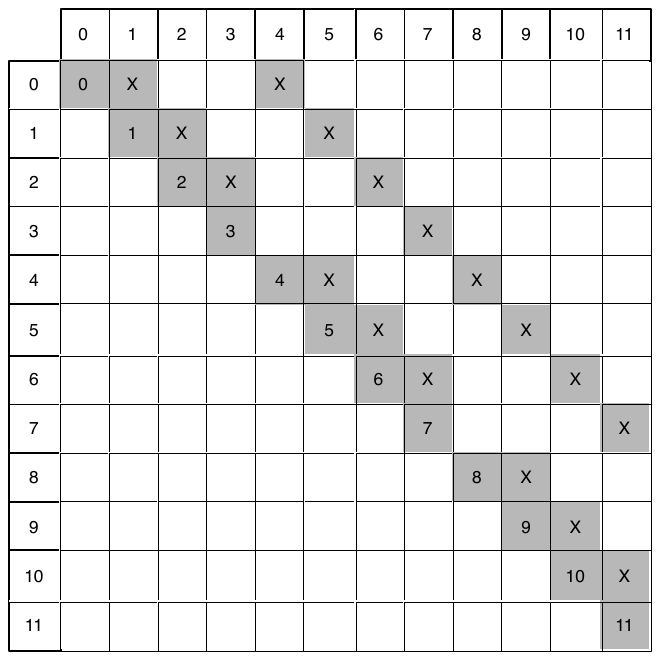}
    } \hspace{0.8cm}
    \subfloat[Block reordering]{
      \includegraphics[scale=\figsize]{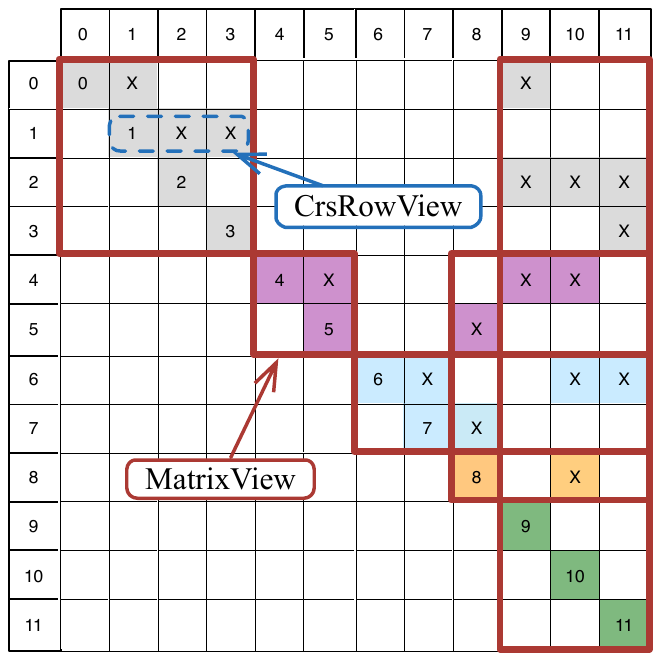}
    }
  \end{center}
  \vspace{\figup}
  \caption{An example of symmetric block nested dissection ordering
    permuted by Scotch. Left: a sparse matrix with natural
    ordering. 
    Right: a hierarchical view of the block
    reordered matrix.}
  \label{fig:example:scotch}
\end{figure}

The factorization algorithm in Tacho is
uniquely characterized by its recursive definition of the sparse matrix
structure. In this approach, a 2D sparse matrix consists of
submatrices 
to define computational blocks on a scalar sparse matrix. 
As a result, our
task-parallel Cholesky factorization has the same look-and-feel
as the scalar Cholesky factorization, greatly improving
programmability. We demonstrate this later in \Sec{algo:byblocks}.

Scotch provides an array of ranges for columns (or rows) in the
reordered matrix where a range corresponds to a group of variables
(separator) that can be treated together. 
Based on the hierarchical relation of the ranges, we
can construct a 2D sparse block layout over the scalar matrix by creating
view objects that cover nonzero regions. For example,
\Fig{example:scotch} illustrates the reordered sparse matrix and its
corresponding block structure. We denote this collection of submatrices as
a 2D sparse partitioned-block matrix. 



To facilitate the 2D block layout, we propose the following hierarchy of views on a
sparse matrix:
\begin{description}
\item[CrsMatrixBase]
  a base matrix object that contains the standard data structure for
  sparse matrices \ie, row-pointers, column-index array, and value
  array; 
\item[MatrixView] 
  a matrix view that defines a 2D rectangular data region overlaid on the
  base matrix, which is defined by offsets and view dimensions, see \Fig{example:scotch};
\item[CrsRowView]
  a sparse row view that defines the range of columns of a row associated
  with a \verb+MatrixView+;  
\item[TaskView]
  a derived class extended from the \verb+MatrixView+ to
  include a future associated with a corresponding 2D data region. 
\end{description}
We assume that matrices are stored in \verb+CrsMatrixBase+ using 
the standard \acf{CSR} format. This base matrix has template arguments
for a value type which can be either a scalar (for a scalar matrix) or
sparse block (for a 2D matrix).
A light-weight matrix view is defined as \verb+MatrixView+ with
partition information. As the matrix view is templated with an associated base
matrix, it could be a block of scalars or block of 2D
matrices. This view object becomes a basic computing unit in our
task-parallel sparse matrix factorization. Task granularity is controlled by
adjusting the size of a matrix view; a view can be split into many
views or views can be merged into a single view. Since the matrix
view only contains meta data, these operations do not carry overhead
of data repacking. Our current work does not include
precise blocksize tuning capabilities for generating optimal task
granularity. Instead, we roughly control the task granularity by
adjusting the Scotch tree hierarchy level.

In addition, an extension of
the matrix view is used for the tasking interface, called \verb+TaskView+.
This class contains a
\verb+future+ object to record a future state updated by tasks
associated with a particular 2D block. Finally, the \verb+CrsRowView+ 
specifies a part of
a row within a matrix view. The row view is used to access elements
of a matrix view in which each element can be either a scalar or sparse 
block matrix itself. 

\paragraph{Difference from other approaches.}
After we construct a 2D matrix based on
\ac{ND} block ordering, we no longer use the tree hierarchy to extract
parallelism. Tasks are created by the algorithms-by-blocks based on the
2D block sparse layout.
The approach differs from others in that we do not explicitly rely on
the \ac{ND} tree-hierarchy (or the elimination tree) in the numeric factorization
phase. On the other hand, conventional approaches for task-parallel
implementation explicitly use the tree-hierarchy to generate independent
tasks and their dependences, as well as to distribute compute
resources according to subtree structures~\cite{Aliaga:2014:taskilu}.
Such implementations 
may not be performance-portable as the implementation is
hard-wired to problem-specific sparse structures and 
hardware execution environments.




\subsection{Numeric factorization: Cholesky-by-blocks}
\label{sec:algo:byblocks}

\begin{figure}[tb!]
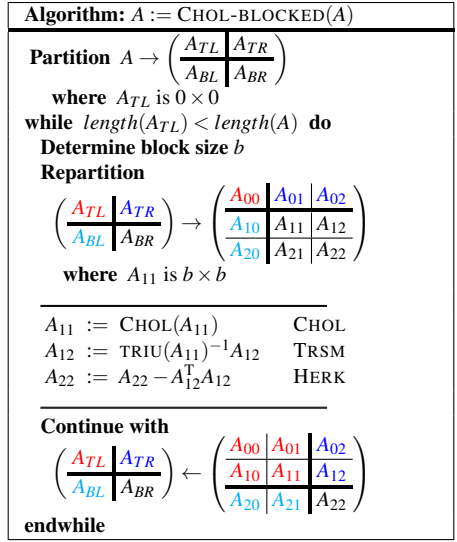

  \begin{center}     
      \resetsteps      


      \renewcommand{\operation}{  A \becomes \mbox{\sc Chol-blocked}( A ) }
      \renewcommand{\routinename}{ A \becomes \mbox{\sc Chol-blocked}( A ) }


      \renewcommand{\guard}{
        length( A_{TL} ) < length( A )
      }


      \renewcommand{\update}{
        $ 
        \begin{array}{rcll}
    A_{11} &\becomes& \mbox{\sc Chol}( A_{11} ) & \quad\mbox{\sc Chol}  \\
    A_{12} &\becomes& \triu{A_{11}}^{-1} A_{12} & \quad\mbox{\sc Trsm}  \\
    A_{22} &\becomes& A_{22} - \tr{A}_{12} A_{12} & \quad\mbox{\sc Herk}   
        \end{array}
        $
      }


      \renewcommand{\partitionings}{
        $
        A \rightarrow
        \FlaTwoByTwo{A_{TL}}{A_{TR}}
                    {A_{BL}}{A_{BR}}
                    $
      }

      \renewcommand{\partitionsizes}{
        $ A_{TL} $ is $ 0 \times 0 $
      }


      \renewcommand{\blocksize}{b}
      \renewcommand{\repartitionings}{
        $
        \FlaTwoByTwo{\color{red}A_{TL}}{\color{blue}A_{TR}}
                    {\color{cyan}A_{BL}}{A_{BR}}
                    \rightarrow
                    \FlaThreeByThreeBR{\color{red}A_{00}}{\color{blue}A_{01}}{\color{blue}A_{02}}
                                      {\color{cyan}A_{10}}{A_{11}}{A_{12} }
                                      {\color{cyan}A_{20}}{A_{21}}{A_{22}}
                                      $}

      \renewcommand{\repartitionsizes}{
        $ A_{11} $ is $ \blocksize \times \blocksize $
      }


      \renewcommand{\moveboundaries}{
        $
        \FlaTwoByTwo{\color{red}A_{TL}}{\color{blue}A_{TR}}
                    {\color{cyan}A_{BL}}{A_{BR}}
                    \leftarrow
                    \FlaThreeByThreeTL{\color{red}A_{00}}{\color{red}A_{01}}{\color{blue}A_{02}}
                                      {\color{red}A_{10}}{\color{red}A_{11}}{\color{blue}A_{12} }
                                      {\color{cyan}A_{20}}{\color{cyan}A_{21}}{A_{22}}
                                      $}
      \begin{minipage}[t]{2.35in} 
        \setlength{\arraycolsep}{2pt}
        \footnotesize 
        \FlaAlgorithmNarrow
      \end{minipage}
  \end{center} 
  \vspace{\figup}
  
  \caption{Cholesky algorithm. The blocks in the $2\times2$
  and $3\times3$ block matrices that correspond to each other are of 
  the same color. {\sc Chol} and {\sc triu} represent Cholesky
  factorization and the upper triangular part of an input matrix respectively.}
  \label{fig:dense-chol}
\end{figure}


This section describes the Cholesky-by-blocks algorithm.
The right-looking Cholesky algorithm is shown in \Fig{dense-chol}.
The algorithm is expressed with partitioned matrices using \ac{FLAME}
notations~\cite{Quintana:2001:flame,VanDeGeijn:2008:flame}. A short
description of the notation follows. First, note that the
algorithm will work equally well on matrices with scalar entries ($b$=1)
or sparse block entries (variable $b$). Second, $A_{BR}$ is
partitioned further and updated at each iteration of the {\tt while} loop.
In this particular algorithm, note that the computations only happen
on the $A_{BR}$ block. The algorithm
consists of three different operations. Using the \ac{BLAS} notation, the
three operations in \Fig{dense-chol} are {\sc Chol}, {\sc Trsm}, and {\sc Herk} which correspond
to a Cholesky factorization, triangular solve and hermitian rank-k update.
Finally, the partition is redefined (see the thick partition line moving 
forward) for the next iteration of the algorithm.

\renewcommand{\FlaStartComputeShorter}{
  \setlength{\unitlength}{0.5in}
  \begin{picture}(5.5,0.01)
    \put(0,0){\line(1,0){5.5}}
    \put(0,0.01){\line(1,0){5.5}}
  \end{picture}
}  
\renewcommand{\FlaEndComputeShorter}{
  \setlength{\unitlength}{0.5in}
  \begin{picture}(5.5,0.01)
    \put(0,0){\line(1,0){5.5}}
    \put(0,0.01){\line(1,0){5.5}}
  \end{picture}
}  

\begin{figure}[tb!]
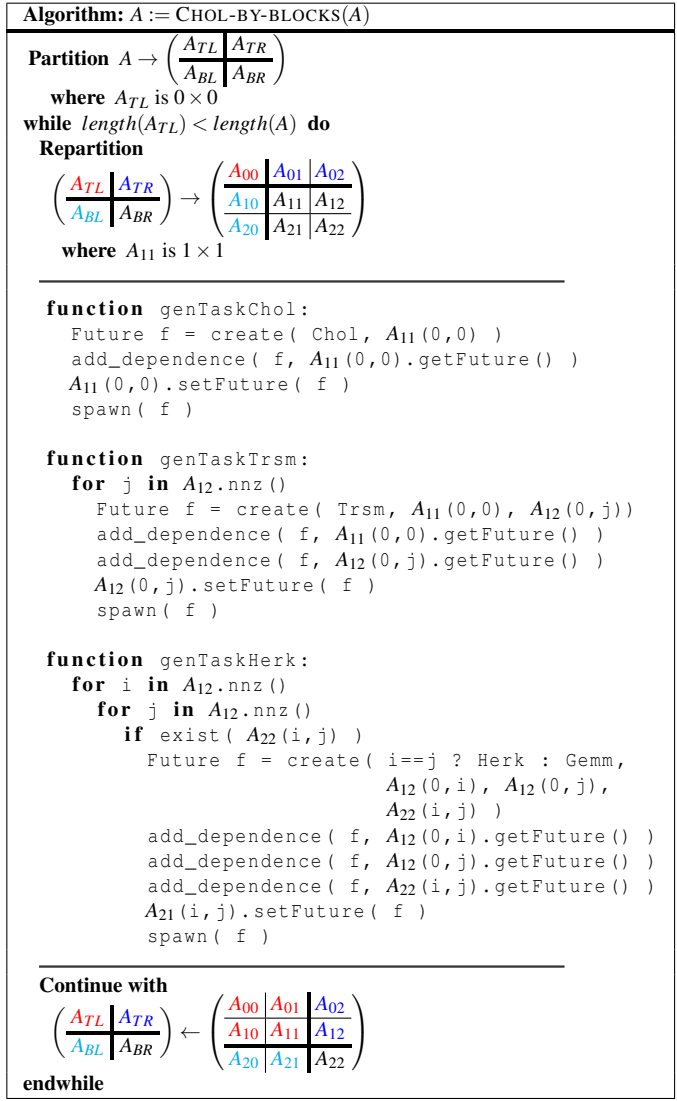

  \begin{center}     
      \resetsteps      


      \renewcommand{\operation}{  A \becomes \mbox{\sc Chol-by-blocks}( A ) }
      \renewcommand{\routinename}{ A \becomes \mbox{\sc Chol-by-blocks}( A ) }


      \renewcommand{\guard}{
        length( A_{TL} ) < length( A )
      }

      
      \renewcommand{\update}{
      }


      \renewcommand{\partitionings}{
        $
        A \rightarrow
        \FlaTwoByTwo{A_{TL}}{A_{TR}}
                    {A_{BL}}{A_{BR}}
                    $
      }

      \renewcommand{\partitionsizes}{
        $ A_{TL} $ is $ 0 \times 0 $
      }


      \renewcommand{\blocksize}{1}
      \renewcommand{\repartitionings}{
        $
        \FlaTwoByTwo{\color{red}A_{TL}}{\color{blue}A_{TR}}
                    {\color{cyan}A_{BL}}{A_{BR}}
                    \rightarrow
                    \FlaThreeByThreeBR{\color{red}A_{00}}{\color{blue}A_{01}}{\color{blue}A_{02}}
                                      {\color{cyan}A_{10}}{A_{11}}{A_{12} }
                                      {\color{cyan}A_{20}}{A_{21}}{A_{22}}
                                      $}

      \renewcommand{\repartitionsizes}{
        $ A_{11} $ is $ \blocksize \times \blocksize $
      }


      \renewcommand{\moveboundaries}{
        $
        \FlaTwoByTwo{\color{red}A_{TL}}{\color{blue}A_{TR}}
                    {\color{cyan}A_{BL}}{A_{BR}}
                    \leftarrow
                    \FlaThreeByThreeTL{\color{red}A_{00}}{\color{red}A_{01}}{\color{blue}A_{02}}
                                      {\color{red}A_{10}}{\color{red}A_{11}}{\color{blue}A_{12} }
                                      {\color{cyan}A_{20}}{\color{cyan}A_{21}}{A_{22}}
                                      $}
      \begin{minipage}[t]{3.2in} 
        \setlength{\arraycolsep}{2pt}
        \footnotesize 
        \begin{center}
          \begin{tabular}{|l |} \hline
            {\bf Algorithm:} $\routinename$ \\ \whline
            {\WSpartitionNarrow} \\
            {\bf while} \WSguard { \bf do} \\
            \ \hspace{0.0in} \WSrepartitionNarrow \\
            {\hspace{0.0in} \FlaStartComputeShorter} \\
            {\hspace{0.0in}
              \lstsetbylang{C++}
              \begin{lstlisting}[mathescape]
function genTaskChol:
  Future f = create( Chol, $A_{11}$(0,0) ) 
  add_dependence( f, $A_{11}$(0,0).getFuture() )
  $A_{11}$(0,0).setFuture( f ) 
  spawn( f ) 

function genTaskTrsm:
  for j in $A_{12}$.nnz() 
    Future f = create( Trsm, $A_{11}$(0,0), $A_{12}$(0,j)) 
    add_dependence( f, $A_{11}$(0,0).getFuture() ) 
    add_dependence( f, $A_{12}$(0,j).getFuture() ) 
    $A_{12}$(0,j).setFuture( f ) 
    spawn( f ) 

function genTaskHerk:
  for i in $A_{12}$.nnz()
    for j in $A_{12}$.nnz()
      if exist( $A_{22}$(i,j) )
        Future f = create( i==j ? Herk : Gemm,
                           $A_{12}$(0,i), $A_{12}$(0,j),
                           $A_{22}$(i,j) ) 
        add_dependence( f, $A_{12}$(0,i).getFuture() )
        add_dependence( f, $A_{12}$(0,j).getFuture() )
        add_dependence( f, $A_{22}$(i,j).getFuture() )      
        $A_{21}$(i,j).setFuture( f )       
        spawn( f )
              \end{lstlisting} 
            } \\
            {\hspace{0.0in} \FlaEndComputeShorter} \\
            {\ \hspace{0.0in} \WSmoveboundaryNarrow} \\
            {{\bf endwhile}} \\ \hline
          \end{tabular}
        \end{center} 
      \end{minipage}
  \end{center} 
  \vspace{\figup}
  
  \caption{Cholesky-by-blocks algorithm on a 2D partitioned-block
    matrix. The blocks in the $2\times2$
    and $3\times3$ block matrices that correspond to each other are of 
    the same color.}
  \label{fig:sparse-chol}
\end{figure}

We transform this algorithm into Cholesky-by-blocks by
converting the basic computing unit from a scalar to a block.
The blocked algorithm described in \Fig{dense-chol} is
applied to $\bfA_{ij}$ elementwise. 
By doing so, the three different
operations inside the while loop becomes three different opportunities
to generate tasks.
\Fig{sparse-chol} describes the Cholesky-by-blocks algorithm
using the Kokkos tasking interface.
By running the Cholesky-by-blocks on a 2D
matrix, tasks are created and spawned with dependences.
A spawned task
is recorded on a {\tt future} of an output matrix view associated with the task.
The dependence for each task is determined by the input/output blocks
used in the task and any futures associated with them.
Since blocks record associated tasks, we do not need to keep track of
the entire task dependences but only follow the loop body of the
algorithm.

We demonstrate this algorithm with a small example matrix.
Suppose that \ac{ND} ordering provides a symmetric
permutation matrix $\bfP$, which leads to an upper triangular block matrix
\[
\bfP^{T}\bfA\bfP = \left(
\begin{array}{ c c c c c }
  A_{00} & \quad  & \quad  & \quad  & A_{04} \\
  \quad  & A_{11} & \quad  & A_{13} & A_{14} \\
  \quad  & \quad  & A_{22} & A_{23} & A_{24} \\
  \quad  & \quad  & \quad  & A_{33} & A_{34} \\
  \quad  & \quad  & \quad  & \quad  & A_{44}
\end{array}
\right)
\]
where all $\bfA_{ij}$ blocks are sparse and have block
dimensions compatible with each other. Then, we apply the Cholesky-by-blocks
algorithm as depicted in \Fig{sparse-chol}.
As a result, a sequence of block matrix computations is generated as
illustrated in \Fig{example:cholesky-by-blocks}.
%
\begin{figure}[tb!]
  {\small 
    \begin{center}
      \begin{tabular}{l}
        \subfloat[1st iteration]{
          $
          \left(
          \begin{array}{ I c | c c c c }  \whline
            A_{00} & \quad  & \quad  & \quad  & A_{04} \\ \hline
            \quad  & A_{11} & \quad  & A_{13} & A_{14} \\ 
            \quad  & \quad  & A_{22} & A_{23} & A_{24} \\
            \quad  & \quad  & \quad  & A_{33} & A_{34} \\
            \quad  & \quad  & \quad  & \quad  & A_{44} \\
          \end{array}
          \right)
          \qquad
          \begin{array}{ rcl }
            A_{00} &\becomes& \Chol{A_{00}} \\
            A_{04} &\becomes& \triu{A_{00}}^{-1} A_{04} \\
            A_{44} &\becomes& A_{44} - A_{04}^{T} A_{04} \\
          \end{array} 
          $
        } \\
        \subfloat[2nd iteration]{
          $
          \left(
          \begin{array}{ c I c | c c c } 
            A_{00} & \quad  & \quad  & \quad  & A_{04} \\ \whline
            \quad  & A_{11} & \quad  & A_{13} & A_{14} \\ \hline
            \quad  & \quad  & A_{22} & A_{23} & A_{24} \\
            \quad  & \quad  & \quad  & A_{33} & A_{34} \\
            \quad  & \quad  & \quad  & \quad  & A_{44} \\
          \end{array}
          \right)
          \qquad
          \begin{array}{ rcl }
            A_{11} &\becomes& \Chol{A_{11}} \\
            A_{13} &\becomes& \triu{A_{11}}^{-1} A_{13} \\
            A_{14} &\becomes& \triu{A_{11}}^{-1} A_{14} \\
            A_{33} &\becomes& A_{33} - A_{13}^{T} A_{13} \\
            A_{34} &\becomes& A_{34} - A_{13}^{T} A_{14} \\
            A_{44} &\becomes& A_{44} - A_{14}^{T} A_{14} \\
          \end{array} 
          $
        } \\
        \subfloat[3rd iteration]{
          $
          \left(
          \begin{array}{ c c I c | c c } 
            A_{00} & \quad  & \quad  & \quad  & A_{04} \\ 
            \quad  & A_{11} & \quad  & A_{13} & A_{14} \\ \whline
            \quad  & \quad  & A_{22} & A_{23} & A_{24} \\ \hline
            \quad  & \quad  & \quad  & A_{33} & A_{34} \\
            \quad  & \quad  & \quad  & \quad  & A_{44} \\
          \end{array}
          \right)
          \qquad
          \begin{array}{ rcl }
            A_{22} &\becomes& \Chol{A_{22}} \\
            A_{23} &\becomes& \triu{A_{22}}^{-1} A_{23} \\
            A_{24} &\becomes& \triu{A_{22}}^{-1} A_{24} \\
            A_{33} &\becomes& A_{33} - A_{23}^{T} A_{23} \\
            A_{34} &\becomes& A_{34} - A_{23}^{T} A_{24} \\
            A_{44} &\becomes& A_{44} - A_{24}^{T} A_{24} \\
          \end{array} 
          $
        } \\
        \subfloat[4th iteration]{ 
          $
          \left(
          \begin{array}{ c c c I c | c } 
            A_{00} & \quad  & \quad  & \quad  & A_{04} \\ 
            \quad  & A_{11} & \quad  & A_{13} & A_{14} \\ 
            \quad  & \quad  & A_{22} & A_{23} & A_{24} \\ \whline
            \quad  & \quad  & \quad  & A_{33} & A_{34} \\ \hline
            \quad  & \quad  & \quad  & \quad  & A_{44} \\
          \end{array}
          \right)
          \qquad
          \begin{array}{ rcl }
            A_{33} &\becomes& \Chol{A_{33}} \\
            A_{34} &\becomes& \triu{A_{33}}^{-1} A_{34} \\
            A_{44} &\becomes& A_{44} - A_{34}^{T} A_{34} \\
          \end{array} 
          $
        } \\
        \subfloat[5th iteration]{
          $
          \left(
          \begin{array}{ c c c c I c | } 
            A_{00} & \quad  & \quad  & \quad  & A_{04} \\ 
            \quad  & A_{11} & \quad  & A_{13} & A_{14} \\ 
            \quad  & \quad  & A_{22} & A_{23} & A_{24} \\ 
            \quad  & \quad  & \quad  & A_{33} & A_{34} \\ \whline
            \quad  & \quad  & \quad  & \quad  & A_{44} \\   \hline
          \end{array}
          \right)
          \qquad
          \begin{array}{ rcl }
            A_{44} &\becomes& \Chol{A_{44}}  \\
          \end{array} 
          $
        }
      \end{tabular}
    \end{center}
  }
  \vspace{\figup}
  \caption{Generated block matrix computations while proceeding
    on Cholesky-by-blocks.}
  \label{fig:example:cholesky-by-blocks}
\end{figure}
\begin{figure}[tb!]
  \begin{center}
    \subfloat[Block ND tree]{
      \includegraphics[scale=0.8]{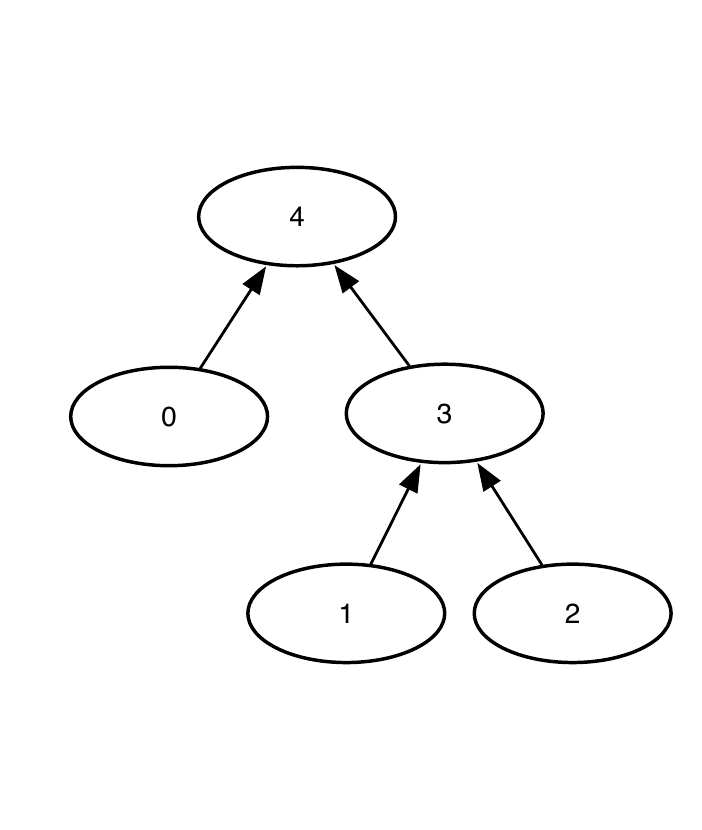}
      \label{fig:example:tasks:nd}
    }
    \subfloat[Task DAG]{
      \includegraphics[scale=0.8]{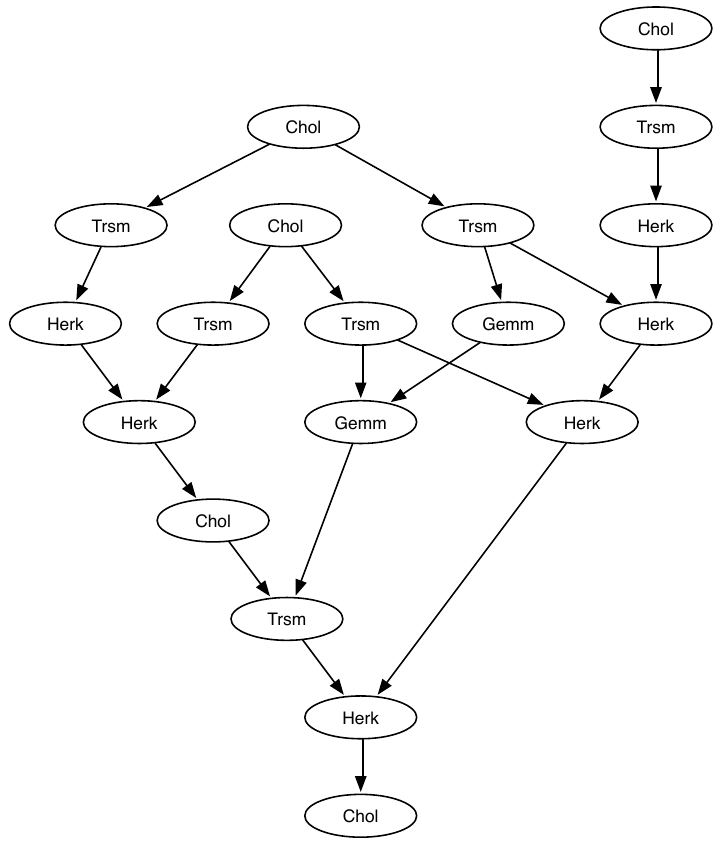}
      \label{fig:example:tasks:dag}
    }
  \end{center}
  \vspace{\figup}
  \caption{A task dependence graph for the example illustrated in
    \Fig{example:cholesky-by-blocks}.}
  \label{fig:example:tasks}
\end{figure}
Task dependences are found from the input/output relations
described in the loop body of the blocked Cholesky algorithm:
\[
\mbox{\sc Chol}\rightarrow\mbox{\sc Trsm}\rightarrow \mbox{{\sc Herk} (or {\sc Gemm})}.
\]
For example, a {\sc Trsm} task will depend on a {\sc Chol} task that
is an input for the task; a {\sc Herk} task will depend on
{\sc Trsm} tasks that should be completed on the data region that
is required as an input of the task.
This dependence relationship is applied to individual
blocks and a corresponding task \ac{DAG} is shown in \Fig{example:tasks:dag}.
Contrary to the coarse grain tasks that correspond to the block \ac{ND}
tree depicted in \Fig{example:tasks:nd}, our sparse Cholesky-by-blocks
generates a larger number of asynchronous fine-grained tasks.
In this particular example, the
$\mbox{\sc Chol}(A_{22})$ in the third iteration can be executed
before finishing tasks (\ie, $\mbox{\sc Trsm}$, $\mbox{\sc Herk}$ and
$\mbox{\sc Gemm}$) created in the first and second iterations.
This is possible because the $\mbox{\sc Chol}(A_{22})$ has input dependences
only to itself. Our Cholesky-by-blocks algorithm does not do any special
book-keeping to determine when to launch a certain task. Instead, it 
loops through the entire 2D matrix 
and generates the tasks as it steps through the loop.
Tasks such as $\mbox{\sc Chol}(A_{22})$
can be run immediately as they are created
with no dependences.
This is possible because our task-parallel approach is not strictly
tied to the tree hierarchy derived from \ac{ND} ordering.
Furthermore, we see in the first iteration that $\mbox{\sc Trsm}(A_{04})$ can begin
immediately as its dependences are satisfied, whereas $\mbox{\sc Chol}(A_{22})$
won't be even created till the second iteration, which is the opposite of what
a tree based algorithm would have done.
This approach exposes much more fine-grained task parallelism. Also
note that $\mbox{\sc Herk}(A_{33})$ in the third iteration has taken
the form of a sparse {\sc Gemm} which is much more 
cache-friendly to compute than simple rank-1 updates.

\section{Performance evaluation}
\label{sec:result}

\begin{table}[tb!]
  \begin{center}
    {\small 
      \begin{tabular}{l r r r}    \toprule
        Matrix ID             & \# of rows(n)  & \# of nonzeros(nnz) & nnz/n   \\\midrule
        \verb+ecology2+       & 999,999        & 4,995,991    & 4.99\\
        \verb+G3_circuit+     & 1,585,478      & 7,660,826    & 4.83\\
        \verb+parabolic_fem+  & 525,825        & 3,674,625    & 6.98\\
        \verb+thermal2+       & 1,228,045      & 8,580,313    & 6.98\\
        \verb+bmwcra_1+       & 148,770        & 10,641,602   & 71.53 \\
        \verb+pwtk+           & 217,918        & 11,524,432   & 52.88 \\\bottomrule
      \end{tabular}
    }
  \end{center}
  \vspace{\figup}
  \caption{Test problems selected from the University of Florida sparse matrix collection. }
  \label{tab:test-problems}
\end{table}

In this section, we evaluate our task-parallel incomplete
Cholesky factorization on problems
selected from the University of
Florida sparse matrix collection~\cite{Davis:2011:matrixcollection}.
The matrix properties are tabulated in \Tab{test-problems}.
The largest problem, \verb+G3_circuit+, has about 1.5 million
rows. Note that \verb+bmwcra_1+ and \verb+pwtk+ are 
relatively denser by
an order of magnitude than other test problems
(see the average number of non-zeros per row in the table). Later, we 
show that this property significantly changes
both serial and parallel performance.
All experiments are performed on 
a machine with a dual socket configuration of ``Sandy Bridge''
processors ($2\times8$ Xeon E5-2670 2.6GHz cores) 
and two ``Knights Corner'' coprocessors ($1\times57$ Xeon Phi with
1.1GHz cores) connected via PCI-Express.  
Each Intel Xeon Phi coprocessor has 57 cores with 4 hyperthreads per
core. 
We turn off hyperthreading using a hardware
locality library~\cite{Broquedis:2010:hwloc} and use up to 56 cores
in a native mode 
as one core is reserved for the operating system. 
The GNU compiler (5.1.0) and Intel compiler (15.2.164) with -O3 and Kokkos Pthreads and
Qthreads backends are used for the Sandy Bridge multicore processor
and Xeon Phi coprocessor respectively.

To benchmark parallel performance, we compare Tacho
to a parallel ILU package,
Euclid~\cite{Hysom:2001:ilu} as the closest alternative. The Euclid 
library is a scalable
implementation of the parallel ILU factorization using \ac{MPI}.
The Euclid parallel ILU algorithm consists of four phases: 1) a partitioning
phase to minimize communication costs, 2) local reordering to separate
interior nodes from boundary nodes, 3) global reordering to improve parallelism
and 4) the numeric factorization.
In this comparison, we use symmetric \ac{RCM} ordering for Euclid to
reduce the bandwidth of matrices as this is the best possible ordering for
Euclid's performance. A matrix is distributed to
processors such that each processor owns an equal number of nodes. As
the two codes has different reordering strategies and symbolic
factorization phases, we compare the numeric factorization phase
of Euclid with our factorization code.
Note that Euclid is an LU factorization code as opposed to Cholesky
factorization that we do here. As a result, we divide Euclid's numbers by
half to approximate the factorization costs. It is important not to place
a huge emphasis on these performance numbers as we are comparing an \ac{MPI} based
code with a shared-memory code. However, this is the closest codebase that is
publicly available
for parallel incomplete factorization.
We present these numbers just to demonstrate the difference
between coarse-grained parallelism with traditional rowwise layouts and
fine-grained parallelism with 2D layouts.


\subsection{Symbolic factorization results}

\begin{table}[tb!]
  \begin{center}
    { \small 
      \begin{tabular}{l r r r r r}    \toprule
        \quad & \multicolumn{5}{c}{\# of nonzeros in $U$[millions]} \\\cmidrule(lr){2-6}
        Matrix ID             & L0 & L1 & L2 & L4 & Chol(AMD) \\\midrule
        \verb+ecology2+       & 2.9 &  4.7 & 6.0 & 8.3 & 45.7 \\
        \verb+G3_circuit+     & 4.6 &  7.4 & 9.6 & 14.5 & 189.1 \\
        \verb+parabolic_fem+  & 2.1 &  3.5 & 4.8 & 6.9 & 36.1 \\
        \verb+thermal2+       & 4.9 &  7.9 & 10.6 & 14.8 & 64.8    \\\midrule
        \verb+bmwcra_1+       & 5.3 & 14.1 & 23.2 & 38.5 & 90.9 \\
        \verb+pwtk+           & 5.9 & 10.9 & 15.0 & 21.4 & 60.0 \\\bottomrule
      \end{tabular}
    }
  \end{center}
  \vspace{\figup}
  \caption{Number of nonzero $U$ factors resulting from level(k)
    symbolic factorization; for comparison, the last column shows the number of
    fill from complete factorization with AMD ordering.}
  \label{tab:test-problems:fill}
\end{table}

The symbolic factorization phase determines the location of fill 
for the level(k) incomplete Cholesky factorization. 
Similar to Hysom and Pothen~\cite{Hysom:2002:ilu}, our
symbolic factorization performs \ac{BFS} on the adjacency graph 
of a matrix in parallel for each node to specify level(k) fill
structure. This is implemented in a scalable fashion using two Kokkos 
parallel patterns: \verb+parallel_for+ and \verb+parallel_scan+.
The numbers of nonzero $U$ factors generated by the level(k)
incomplete Cholesky factorization are summarized in
\Tab{test-problems:fill}. All matrices are reordered with the block
\ac{ND} algorithm provided by the Scotch
library~\cite{scotch:6.0:manual}. 
For comparison, we also provide size of the fill for
complete Cholesky factorization in the last column of the table. We do
not report the times for symbolic factorization as they are not significant.



\subsection{Numeric factorization results}


\paragraph{Parallel performance.}

\renewcommand{\graphwidth}{2.4in}
\renewcommand{\graphheight}{2.0in}

\renewcommand{\legendsize}{0.3in}
\renewcommand{\legendfont}{\scriptsize}
\renewcommand{\axisfont}{\footnotesize}

\begin{figure*}[tb!]
  \begin{center}
    \subfloat{
      \includegraphics[scale=1.0]{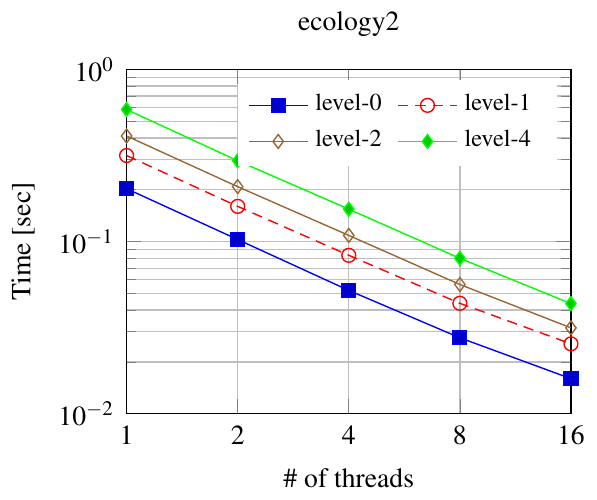}
    } 
    \subfloat{
      \includegraphics[scale=1.0]{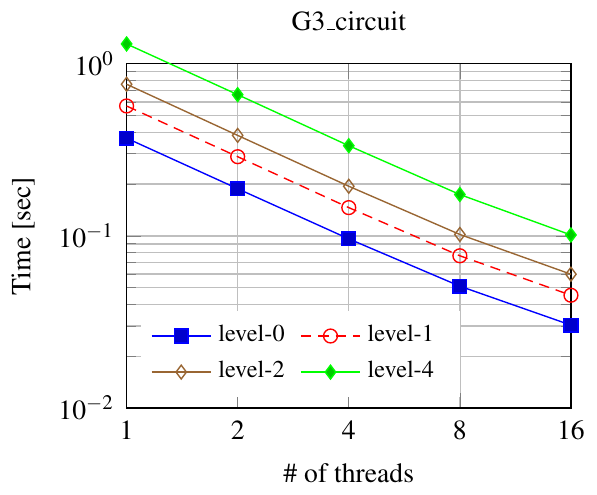}
    } \\
    \subfloat{
      \includegraphics[scale=1.0]{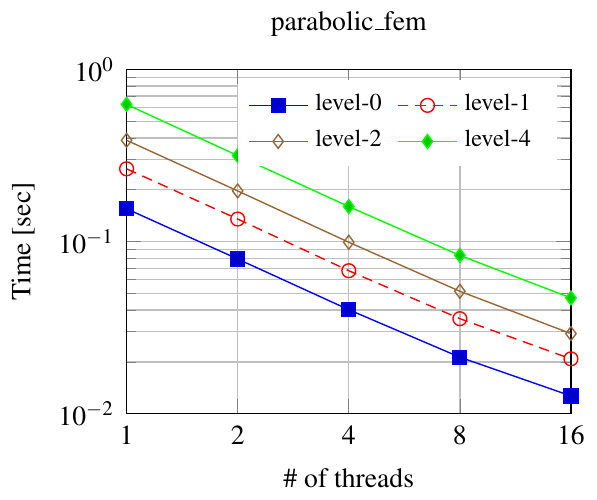}
    }
    \subfloat{
      \includegraphics[scale=1.0]{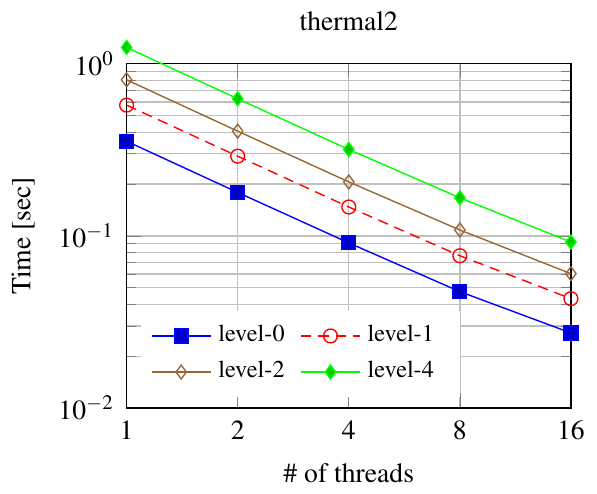}
    } \\ 
    \subfloat{
      \includegraphics[scale=1.0]{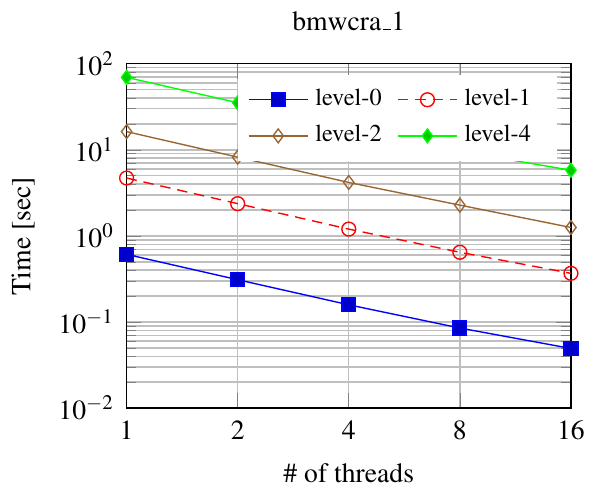}
    }
    \subfloat{
      \includegraphics[scale=1.0]{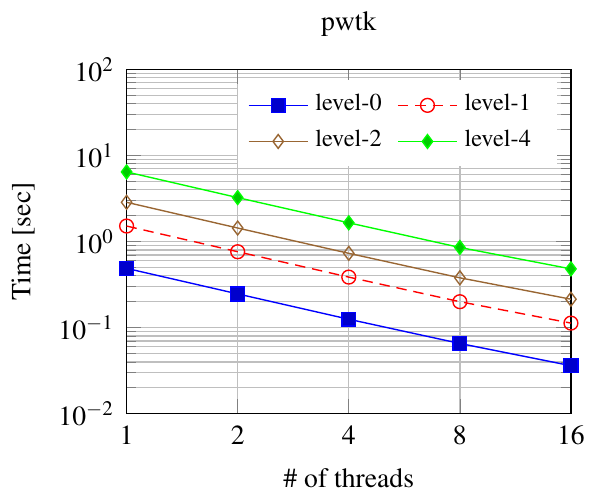}
    } 
  \end{center}
  \caption{[Sandybridge] Time for level(k) incomplete
    Cholesky-by-blocks factorization with the Kokkos Pthreads backend
    (our method).
  } 
  \label{fig:compton-sandybridge:ichol:factorization}
\end{figure*}

\renewcommand{\graphwidth}{2.4in}
\renewcommand{\graphheight}{2.0in}

\renewcommand{\legendsize}{0.3in}
\renewcommand{\legendfont}{\scriptsize}
\renewcommand{\axisfont}{\footnotesize}

\begin{figure*}[tb!]
  \begin{center}
    \subfloat{
      \includegraphics[scale=1.0]{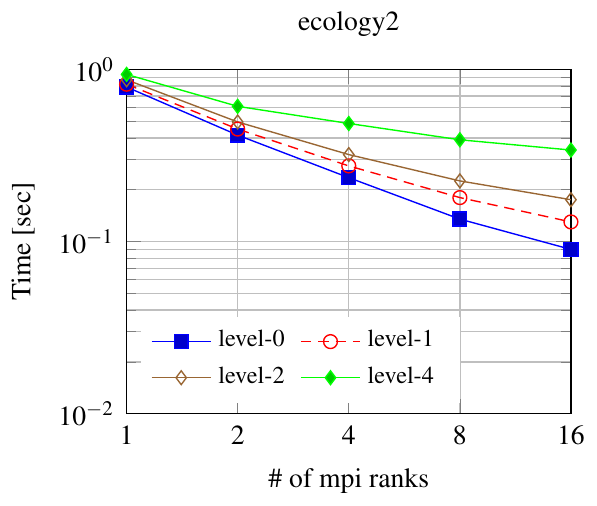}
    } 
    \subfloat{
      \includegraphics[scale=1.0]{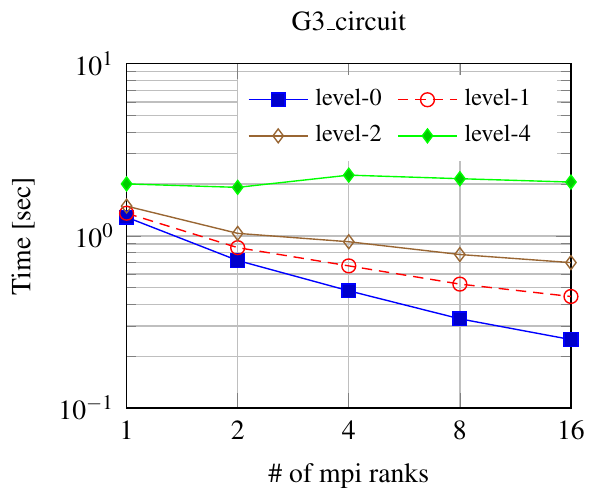}
    } \\
    \subfloat{
      \includegraphics[scale=1.0]{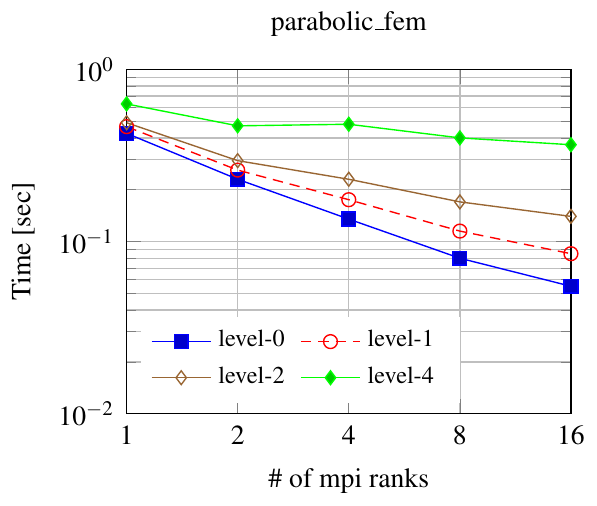}
    }
    \subfloat{
      \includegraphics[scale=1.0]{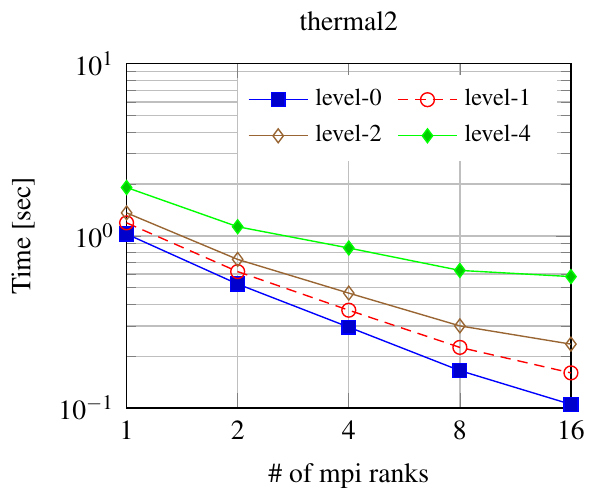}
    } \\
    \subfloat{
      \includegraphics[scale=1.0]{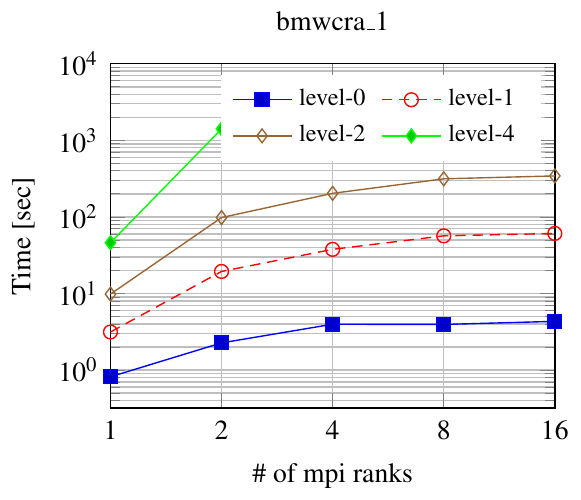}
    }
    \subfloat{
      \includegraphics[scale=1.0]{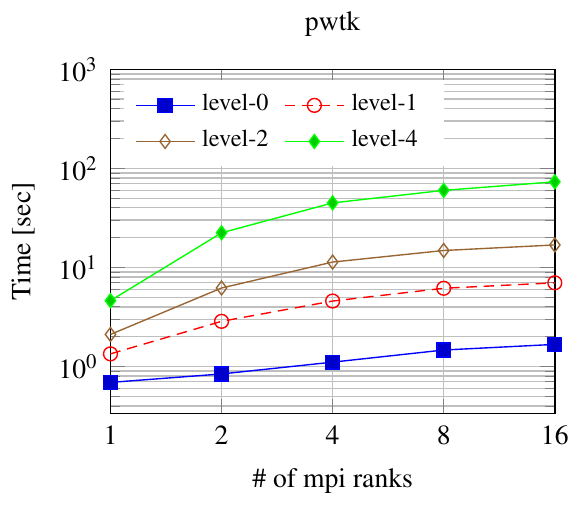}
    } 
  \end{center}
  \caption{[Sandybridge] Time for Euclid level(k) incomplete LU
    factorization (for comparison). The time cost is divided by two to
    compare with the result of incomplete Cholesky factorization. 
    Note that some plots have
    different ranges of values from the plots drawn in 
    \Fig{compton-sandybridge:ichol:factorization}. }
  \label{fig:compton-sandybridge:hypre:factorization}
\end{figure*}

We report strong scalability of our task-parallel level(k) incomplete
Cholesky factorization and evaluate the parallel performance against
the Euclid package. 
Since both Kokkos Pthreads and Qthreads backends report similar timing
results, here we report only for Kokkos Pthreads results. 
First, we compare the time taken for factorization on the Intel
Xeon multicore architecture.
\Fig{compton-sandybridge:ichol:factorization} and
\Fig{compton-sandybridge:hypre:factorization} show the numeric factorization
time for our task-parallel Cholesky
factorization and Euclid respectively. 
Euclid does not provide separate timing
results for symbolic factorization.\footnote{Euclid reports timing
  results for subdomain graph setup, factorization, and solve setup.}
As the symbolic factorization costs much less than the
numeric factorization, we directly compare the numeric factorization time of 
our code to the time reported by the factorization phase of Euclid.
As two codes report different ranges of timing results, we cannot plot
all graphs in the same scale and some graphs are plotted with a
different time scale.
While Euclid scales well for some matrices, relatively
denser problems such as \verb+bmwcra_1+ and \verb+pwtk+ prove harder to solve 
with Euclid-like algorithms.
We conjecture that the main reason that these matrices are harder to solve is the
1D rowwise matrix partitions. 
Although the 1D panels are globally reordered to increase 
parallelism, matrices with a large bandwidth (a higher
number of nonzeros per row) are not ideal for such parallelism.
The same performance trend is commonly observed on other test
problems with an increasing level of fills.
The increased number of fills incurs more synchronization bottlenecks
among 1D panels 
and results in the loss of concurrency.
On the other hand, our task-parallel Cholesky factorization delivers
robust parallel scalability for all test problems, as tasks are generated
based on 2D block matrices and executed asynchronously. 

\renewcommand{\graphwidth}{2.4in}
\renewcommand{\graphheight}{2.0in}

\renewcommand{\legendsize}{0.3in}
\renewcommand{\legendfont}{\scriptsize}
\renewcommand{\axisfont}{\footnotesize}

\begin{figure*}[tb!]
  \begin{center}
    \subfloat{
      \includegraphics[scale=1.0]{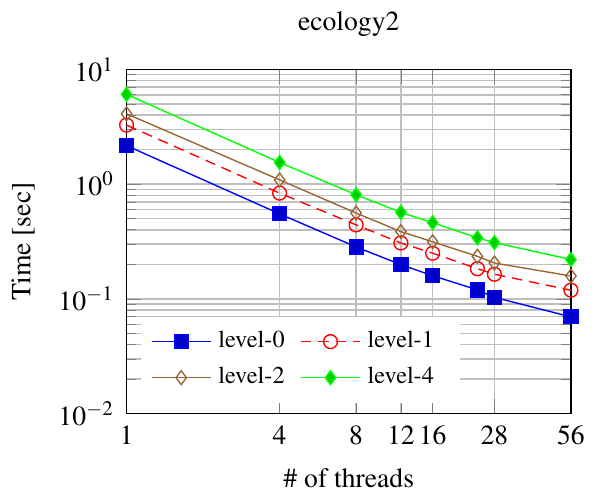}
    }
    \subfloat{
      \includegraphics[scale=1.0]{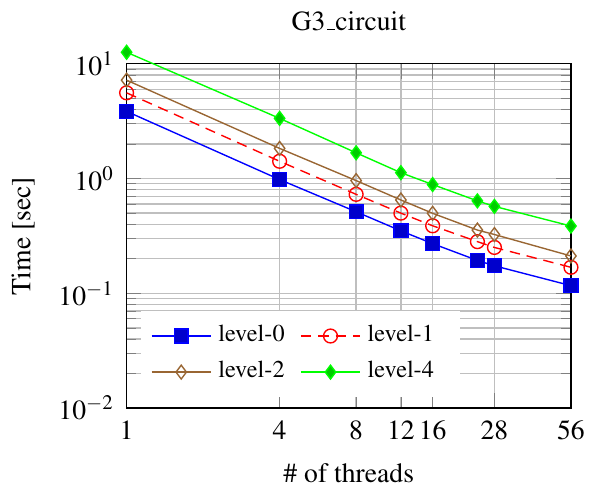}      
    } \\
    \subfloat{
      \includegraphics[scale=1.0]{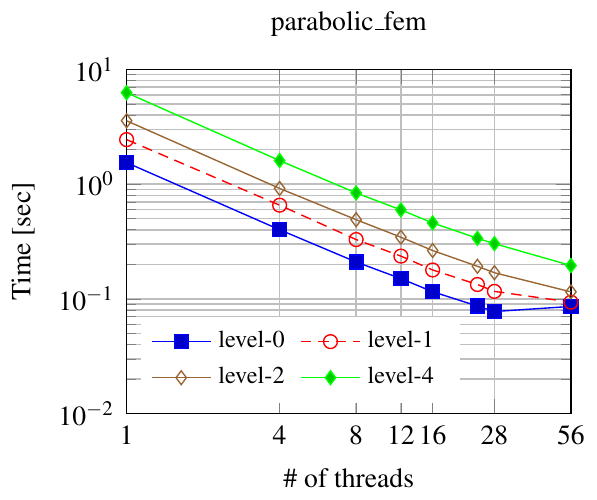}      
    }
    \subfloat{
      \includegraphics[scale=1.0]{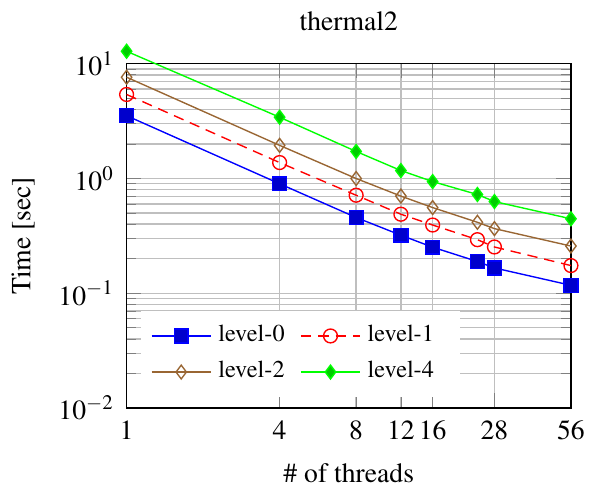}      
    } \\ 
    \subfloat{
      \includegraphics[scale=1.0]{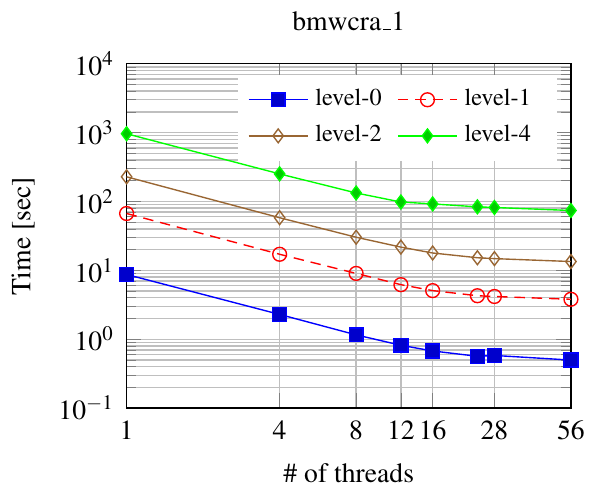}      
    }
    \subfloat{
      \includegraphics[scale=1.0]{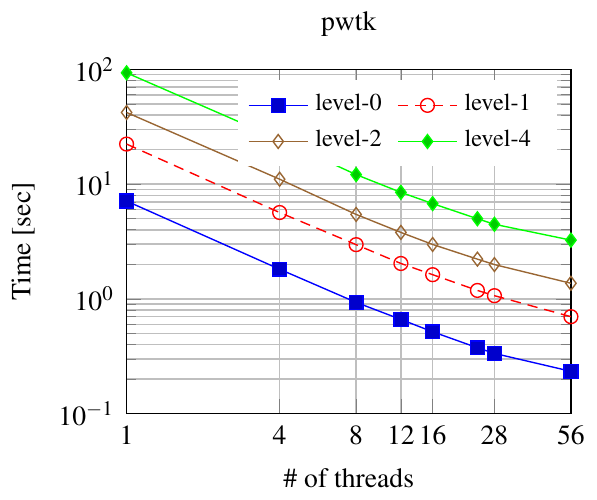}      
    } 

  \end{center}
  \caption{[Phi] Time for level(k) incomplete Cholesky-by-blocks
    factorization with the Kokkos Pthreads backend (our method).}
  \label{fig:compton-phi:ichol:factorization}
\end{figure*}

Comparable parallel performance of our task-parallel Cholesky
factorization on the Intel Xeon Phi coprocessor is illustrated in
\Fig{compton-phi:ichol:factorization}. 
For most test problems, our task-parallel Cholesky algorithm 
scales 
up to the largest number of available
threads on the coprocessor.
Our task-parallel implementation delivers about 26.6x speedup (geometric
mean) over single-threaded Cholesky-by-blocks and 19.2x speedup over serial
Cholesky factorization (which does not carry tasking overhead) using
56 threads on the Intel Xeon Phi processor.


\paragraph{Comparison of tasking overhead between Pthreads and Qthreads.}

\renewcommand{\graphwidth}{4.6in}
\renewcommand{\graphheight}{1.8in}

\renewcommand{\legendsize}{0.3in}
\renewcommand{\legendfont}{\tiny}
\renewcommand{\axisfont}{\footnotesize}

\begin{figure*}[tb!]
  \begin{center}

    \subfloat{
      \includegraphics[scale=1.0]{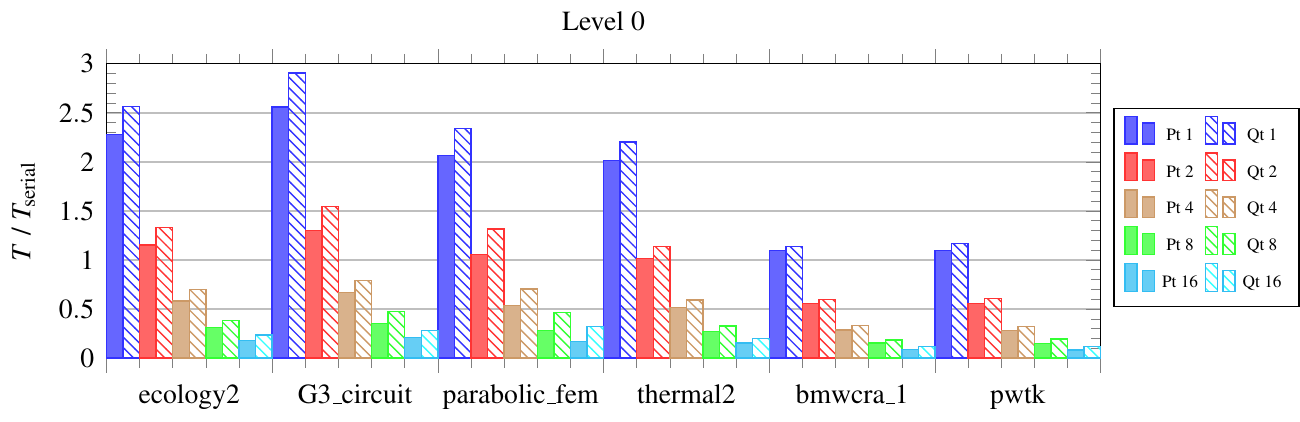} 
    }\\\vspace{-.25cm}
    \subfloat{
      \includegraphics[scale=1.0]{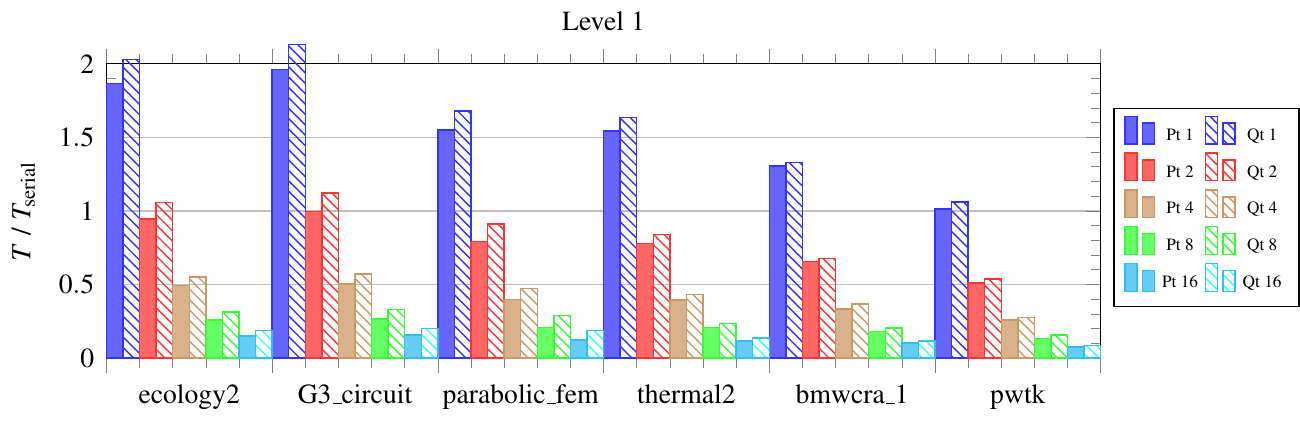} 
    } \\\vspace{-.25cm}
    \subfloat{
      \includegraphics[scale=1.0]{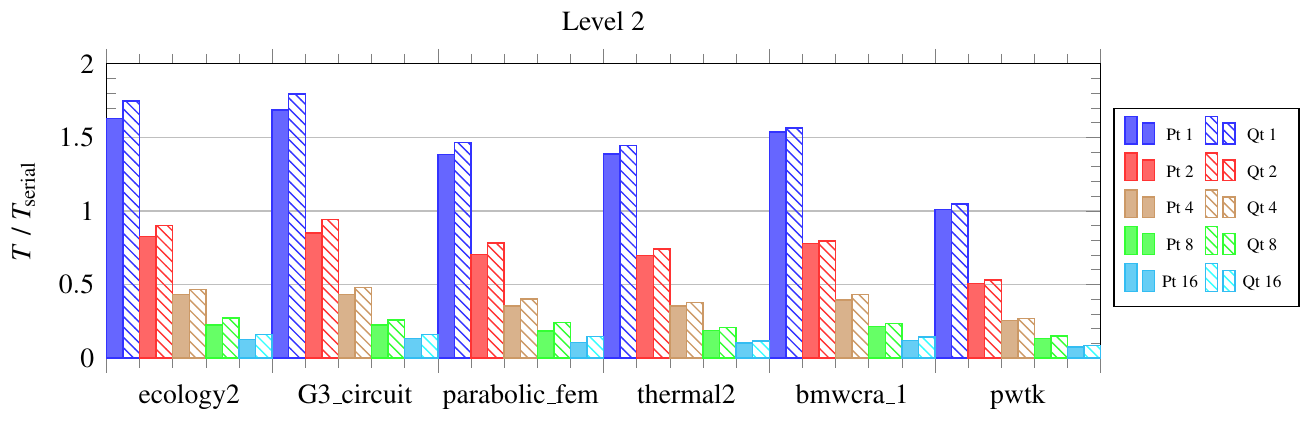} 
    } \\\vspace{-.25cm}
    \subfloat{
      \includegraphics[scale=1.0]{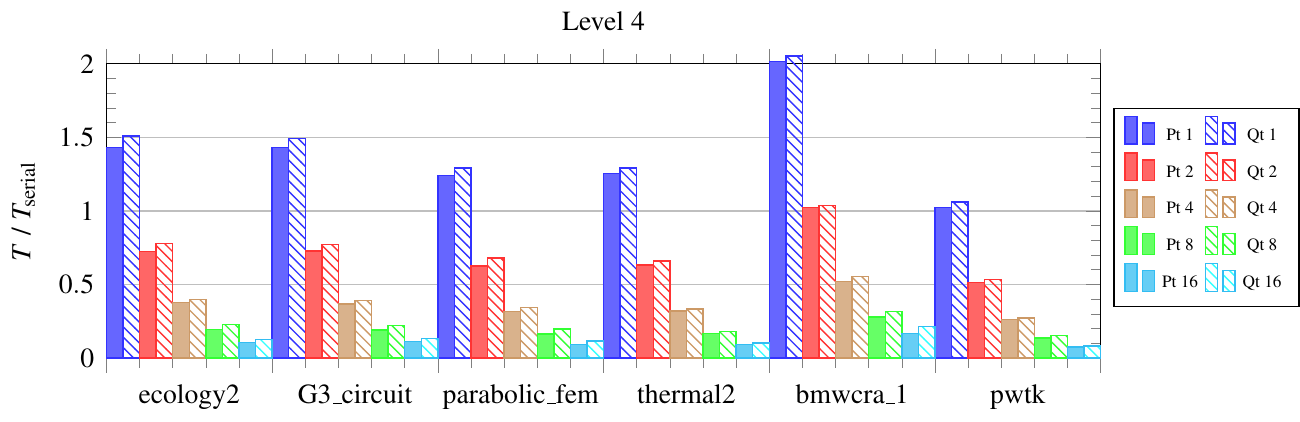} 
    }

  \end{center}
  \caption{[Sandybridge] Time ratio between threaded incomplete
    Cholesky-by-blocks and serial version of incomplete Cholesky
    factorization.}
  \label{fig:compton-sandybridge:ichol:factorization:pq}
\end{figure*}

\renewcommand{\graphwidth}{4.6in}
\renewcommand{\graphheight}{1.8in}

\renewcommand{\legendsize}{0.3in}
\renewcommand{\legendfont}{\tiny}
\renewcommand{\axisfont}{\footnotesize}

\begin{figure*}[tb!]
  \begin{center}

    \subfloat{
      \includegraphics[scale=1.0]{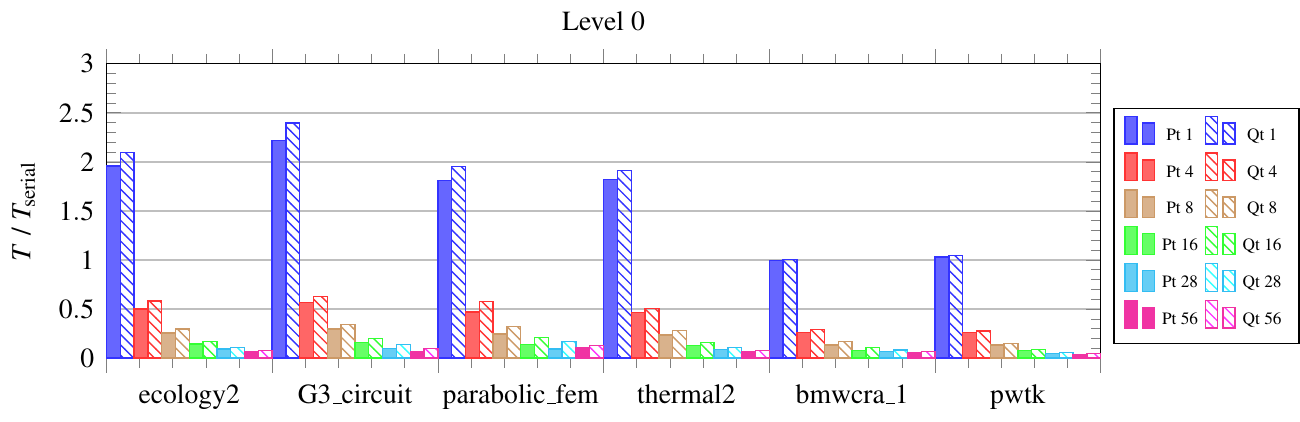} 
    }\\\vspace{-.25cm}    
    \subfloat{
      \includegraphics[scale=1.0]{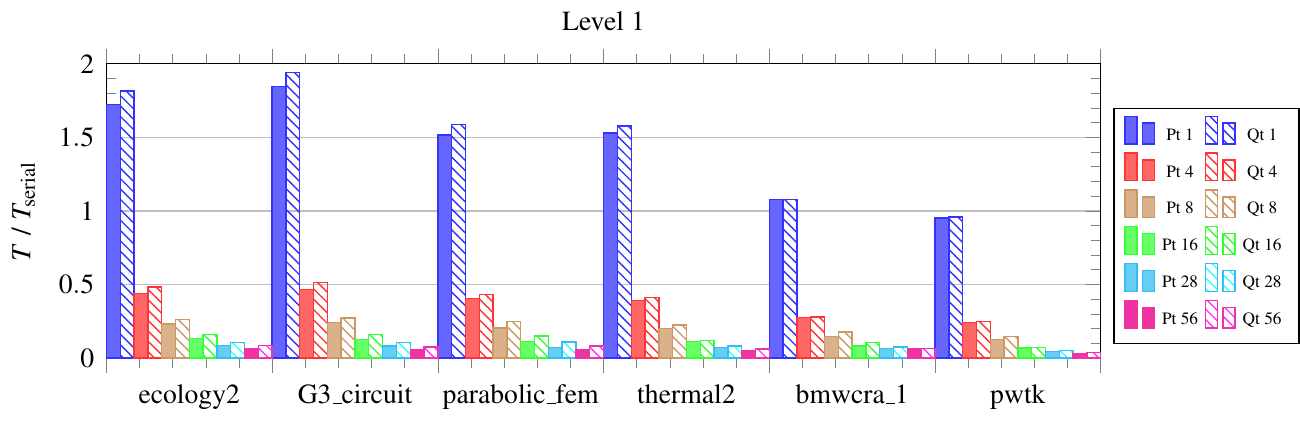} 
    } \\\vspace{-.25cm}
    \subfloat{
      \includegraphics[scale=1.0]{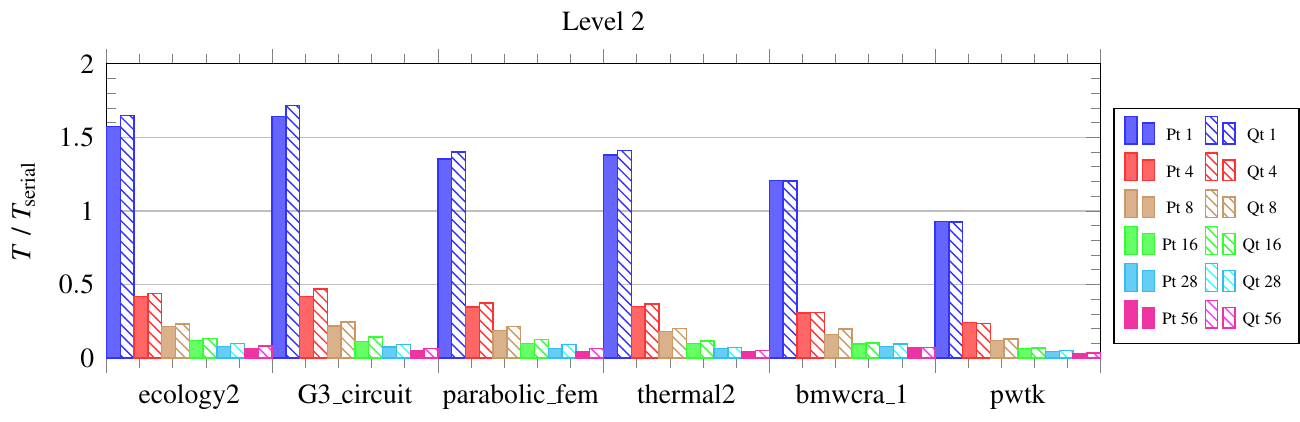} 
    } \\\vspace{-.25cm}
    \subfloat{
      \includegraphics[scale=1.0]{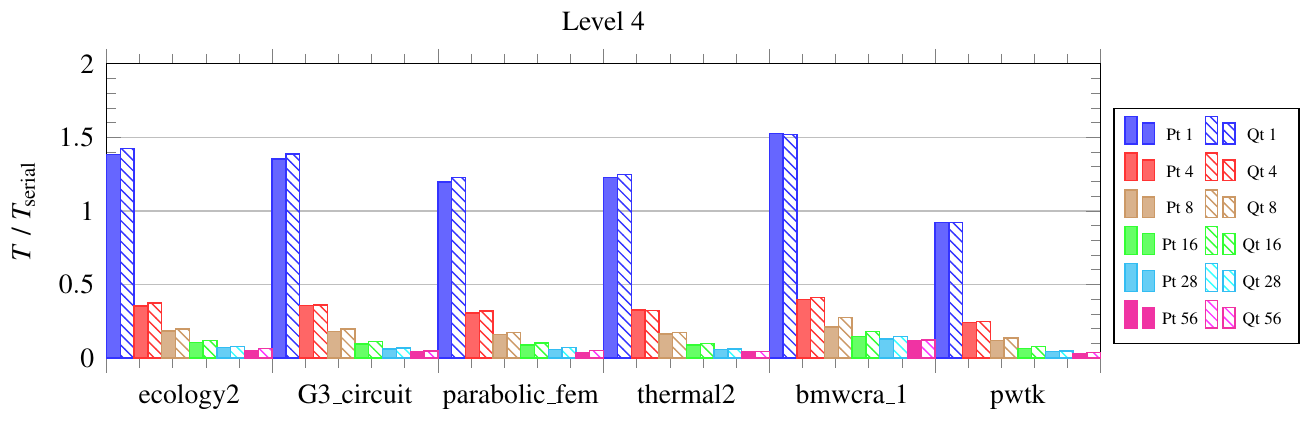} 
    }

  \end{center}
  \caption{[Phi] Time ratio between threaded incomplete
    Cholesky-by-blocks and serial version of incomplete Cholesky
    factorization.}
  \label{fig:compton-phi:ichol:factorization:pq}
\end{figure*}

\begin{table}[tb!]
  \begin{center}
    {\small 
      \begin{tabular}{l r r r}    \toprule
        Matrix ID             & prune level & \# of ranges  & \# of blocks   \\\midrule
        \verb+ecology2+       & 10 & 356      & 1,050   \\
        \verb+G3_circuit+     & 10 & 572      & 2,082   \\
        \verb+parabolic_fem+  & 8  & 668      & 2,437   \\
        \verb+thermal2+       & 10 & 378      & 1,346   \\
        \verb+bmwcra_1+       & 4  & 470      & 1,953   \\
        \verb+pwtk+           & 4  & 778      & 2,933   \\\bottomrule
      \end{tabular}
    }
  \end{center}
  \vspace{\figup}
  \caption{Structure of 2D sparse partitioned-block matrices.
    The prune level lists the height of pruned ND subtrees from the
    leaf level.
    The number of ranges is the total number of vertices in
    a ND tree. The number of blocks is the number of blocks in a 2D matrix.} 
  \label{tab:test-problems:hier}
\end{table}




Finding appropriate task granularity is very important to attain 
higher parallel performance. Multiple aspects of performance
trade-offs should be considered to determine optimal task granularity:
\begin{itemize}
\item total number of generated tasks,
\item level of concurrency expressed from sparse factorization,
\item tasking overhead (context switching, task
  creation, scheduling and destruction),
\item data access overhead for multiple sparse kernel launching,
\item number of computing units and local cache sizes.
\end{itemize}
Using many fine-grained tasks results in a higher degree of concurrency
which is more suitable for manycore computing environments. However, using such a large
number of tasks may significantly increase tasking overhead and
irregular data access cost, which decreases overall
parallel performance of sparse factorization.
On the other hand, generating coarse grained tasks
can decrease tasking overhead 
but may not expose enough concurrency to use all available hardware resources.

To explore this performance trade-off, we plot the relative tasking
overhead, $T/T_\text{serial}$, where $T$ is the time cost of task-parallel
Cholesky-by-blocks and $T_\text{serial}$ is the time cost of
serial sparse Cholesky factorization. 
\Fig{compton-sandybridge:ichol:factorization:pq} describes
the relative tasking overhead on the Sandy Bridge multicore processor
and the Xeon Phi manycore coprocessor.
For the single-threaded case, this measure indicates that if the time
ratio becomes close to one, our Cholesky-by-blocks runs with
relatively small tasking overhead compared to the serial
algorithm. The Cholesky-by-blocks factorization may include
two different types of overhead: 1) task scheduling overhead that
increases proportionally with the number of generated tasks and 2)
overhead due to irregular data access during the asynchronous task
execution. The tasking overhead from our Cholesky-by-blocks can be amortized by
overlapping it during asynchronous task execution.

From the figures, those relatively less sparse matrices such as
\verb+bmwcra_1+ and \verb+pwtk+ shows different performance trends
from the others. For convenience, we use \verb+ecology2+ to represent
the other sparse matrices and use \verb+pwtk+ to represent the less
sparse matrices. Some key observations are:
\begin{itemize}
\item \verb+ecology2+ exhibits higher tasking
  overhead than \verb+pwtk+ due to its lower computational
  workload in each task (for the level 0 factorization,
  \verb+ecology2+ matrix carries almost the same amount of overhead as
  the numerical factorization while the tasking overhead in
  \verb+pwtk+ is almost negligible);   
\item with an increasing level of fill, relative tasking overhead of both
  test problems decreases as the workload associated with each task
  increases;
\item the overhead is problem-specific; for example, the overhead of 
  irregular data access patterns is more dominant for
  \verb+bmwcra_1+, which may results in the increasing overhead with
  the factorization level.
\end{itemize}
The other test problems demonstrate similar
performance behaviors to these two representative cases. 

As depicted in \Fig{compton-phi:ichol:factorization:pq},
similar performance trends are observed on the Xeon Phi manycore
coprocessor. However, the results are quantitatively very distinct
from those obtained on the Sandy Bridge multicore architecture.
The relative time ratio of the single-threaded Cholesky-by-blocks on
the \verb+pwtk+ problem is smaller than one with an increased level of
fill, which implies that the serial factorization that does not
carries tasking overhead is slower than the single-threaded
Cholesky-by-blocks. This counter-intuitive result is probably due
to cache effects.
Since \verb+pwtk+ is considerably less sparse than
\verb+ecology2+, the serial algorithm on this matrix can incur
more cache misses as the effective working set size becomes the
entire matrix. On the other hand, our Cholesky-by-blocks
processes the factorization in terms of block computations. By doing
so, we can effectively reduce cache misses and improve factorization
performance similar to the \ac{BLAS} level 3 operations in
\ac{DLA}.
Consequently, the algorithm designed for task parallelism
is also beneficial for modern manycore architectures
by restricting the computation within a block.
Also, the performance of the serial algorithm on the
multicore architecture is less influenced by the increased amount of
nonzeros because of the large shared L3 cache (20 MB).


\section{Related work}
\label{sec:related-work}

We summarize other task-parallel implementations of sparse direct and
incomplete factorizations as well as other task-parallel models and 
run time systems.

\paragraph{Task-parallel sparse factorization.}

Task-parallel sparse factorization has been implemented mostly
along with multifrontal algorithms, as the algorithm is naturally
parallelized using an elimination tree. This tree-level parallelism
can be easily implemented using tasking \acp{API}. However, the
tree-level parallelism decreases near the root of the tree.
To remedy this inefficiency, nested parallelism within supernodal
blocks is implemented for sparse multifrontal
Cholesky~\cite{Hogg:2010:dag,Irony:2004:sparse_chol},
LU~\cite{Kim:2014:sparse} and
QR~\cite{Buttari:2011:spqr,Davis:2011:spqr}
factorizations.
For an iterative method, ILUPACK~\cite{Aliaga:2014:taskilu} uses a
runtime task scheduler, {\tt OmpSs}, for \ac{PCG}
algorithms. Their approach is similar to ours in that the parallelism
is extracted from the \ac{ND} ordering. However, the code uses
a data-flow programming model, tasks are created using an
elimination tree, and depedences are made through a row (contiguous
memory region).
Hence, their parallel tasks are very fine-grained task operations such
as {\sc Dot} and {\sc Axpy}. By contrast,
we use {\tt future} as a task handle and dependences are made among
future references associated with 2D partitioned-blocks (non
contiguous memory region).
Correct task dependences are derived from algorithms-by-blocks and
tasks are generated separately from the elimination tree, 
enabling a more flexible tasking algorithm. With its 2D
partitioned-block layout, our task-parallel Cholesky factorization
exploits efficient sparse operations \ie, {\sc Chol}, {\sc Trsm},
{\sc Herk} and {\sc Gemm}, which are analogous to \ac{BLAS} level 3
operations in \ac{DLA} libraries.

\paragraph{Task-parallel models and run time systems.}

The need to exploit ever-increasing parallelism on emerging multicore
 and manycore architectures has motivated the development of numerous
task-parallel languages, libraries, and run time systems.
Our tasking model developed in the Kokkos framework
supports both futures and dependences, allowing a
large space of possible task \acp{DAG}, and its \ac{API} and implementation use 
standard C++ with no specialized compiler support needed.
By comparison, Cilk~\cite{cilk} and its successor Intel Cilk
Plus~\footnote{https://www.cilkplus.org} support only strict fork-join
task \acp{DAG} with no dependences or futures, 
though an extension for arbitrary dependences has been explored~\cite{CilkDep}.
OpenMP 4.0~\cite{openmp4} supports
dependences between tasks but does not support futures.
Cilk, Cilk Plus, and OpenMP all require language extensions, and thus, 
special compiler support.
Beginning in version 4.0, Intel \ac{TBB}~\footnote{https://www.threadingbuildingblocks.org} includes a
flow graph interface to represent ``functional nodes'' and edges
between them.
StarPU~\cite{Augonnet:2009:starpu} is a C-based task-parallel framework for 
heterogeneous node architectures with tagged dependences that requires 
extensions to the C language through a GCC plug-in. 
\ac{HPX}~\cite{hpx} supports futures 
across distributed memory machines, 
as do the Chapel~\cite{Chapel} language
and Java-based X10~\cite{x10}, 
and the related Habanero C and Habanero Java~\cite{habaneroJava}. 
Although we use Kokkos to implement our algorithm, 
the techniques we use could be ported to these or similar programming models
with support for futures and dependence-driven execution.

For \ac{DLA}, several research projects
have developed domain-specific runtime task schedulers:
\acs{QUARK}~\cite{QUARK:2011:manual} and
SuperMatrix~\cite{Chan:2007:supermatrix} for shared memory
architectures; DPLASMA~\cite{Bosilca:2011:dplasma} and \acs{PaRSEC}~\cite{Bosilca:2013:parsec} 
for distributed memory
architectures. Recently, a distributed task-parallel Cholesky
implementation~\cite{Martin:2014:smpss} has been demonstrated using
the SMPSs~\cite{Perez:2008:smpss} programming model.

\section{Conclusion}
\label{sec:conclusion}

We have presented a novel algorithm for task-parallel incomplete
Cholesky factorization that applies algorithms-by-blocks factorization to 
a 2D block matrix.
We have shown that by encoding the tree hierarchy in the
2D block matrix, the task DAG need not be restricted to a simple ND tree. This
results in a much richer task DAG, leading to better performance. We
believe this algorithm opens up a new direction of research in which other sparse
factorizations such as LU and QR could also gain performance benefits 
by following the same pattern used here.
We have also designed a simple tasking \ac{API} and modified an open source
library to support task parallelism with performance portable abstractions
for heterogeneous computing devices using different backend
libraries. While used for incomplete Cholesky here, these
changes are much more general and we believe 
they will be useful to develop other task-parallel codes.
We have also shown the performance of a task-parallel Pthreads-based backend
with the incomplete Cholesky factorization as its driver. 
Our factorization has demonstrated robust parallel performance with
several test problems both on
Intel Xeon multicore and Intel Xeon Phi manycore architectures. 
We also evaluated tasking overhead associated with
different task granularities and showed how the overhead costs impact
parallel performance. Due to its sparse nature, finding an optimal
task granularity is a difficult problem compared to the task-parallel
\ac{DLA} libraries previously researched.
We plan to remedy this granularity problem by
exploiting data parallelism within the tasks in the future.
We plan to provide to extend the algorithm for complete factorizations
and provide interfaces through the Amesos2~\cite{Bavier:2012:amesos2}
package in Trilinos.
%


\bibliographystyle{plain}
\bibliography{article}

\end{document}